\documentclass[twocolumn,prd,aps,superscriptaddress,preprintnumbers,tightenlines
,showpacs,nofootinbib,eqsecnum,amsfonts,floatfix]{revtex4}
\pdfoutput=1

\usepackage{color}
\usepackage{calc}
\usepackage{amsmath,amssymb,graphicx}
\usepackage{tensor}
\usepackage{bm}
\usepackage{times}
\usepackage[varg]{txfonts}
\usepackage{float}
\usepackage{dcolumn}
\usepackage[nolist,nohyperlinks]{acronym}
\usepackage{xspace}
\usepackage[abs]{overpic}
\usepackage{pict2e}
\usepackage{enumitem}

\usepackage[utf8]{inputenc}

\usepackage[caption=false]{subfig} 

\DeclareMathAlphabet{\mathcalstd}{OMS}{cmsy}{m}{n}
\DeclareMathAlphabet{\mathpzc}{OT1}{pzc}{m}{it}

\newcommand{\inner}[2]{\left \langle #1 \middle \vert #2 \right \rangle}
\renewcommand{\Re}{\operatorname{Re}}

\newcommand{\Cardiff}{School of Physics and Astronomy, Cardiff University, Queens Building, CF24 3AA, Cardiff, United Kingdom}
\newcommand{\AEI}{Max Planck Institute for Gravitational Physics  (Albert Einstein Institute), Am M\"uhlenberg 1, D-14476 Potsdam-Golm, Germany}

\begin{document}


\title{Can we measure individual black-hole spins from gravitational-wave observations?}

\author{Michael P\"urrer}
\affiliation{\AEI}
\affiliation{\Cardiff}

\author{Mark Hannam}
\affiliation{\Cardiff}

\author{Frank Ohme}
\affiliation{\Cardiff}

\begin{abstract}
Measurements of black-hole spins from gravitational-wave observations of black-hole binaries with ground-based detectors are expected to be hampered by partial degeneracies in the gravitational-wave phasing: between the two component spins, and between the spins and the binary's mass ratio, at least for signals that are dominated by the binary's inspiral. Through the merger and ringdown, however, a different set of degeneracies apply. This suggests the possibility that, if the inspiral, merger and ringdown are all within the sensitive frequency band of a detector, we may be able to break these degeneracies and more accurately measure both spins. In this work we investigate our ability to measure individual spins for non-precessing binaries, for a range of configurations and signal strengths, and conclude that in general the spin of the larger black hole will be measurable (at best) with observations from Advanced LIGO and Virgo. This implies that in many applications waveform models parameterized by only one \emph{effective spin} will be sufficient. Our work does not consider precessing binaries or sub-dominant harmonics, although we provide some arguments why we expect that these will not qualitatively change our conclusions. \textbf{LIGO-P1500255}
\end{abstract}


\maketitle

\acrodef{PN}{post-Newtonian}
\acrodef{GW}{gravitational-wave}
\acrodef{SNR}{signal-to-noise ratio}
\acrodef{aLIGO}{Advanced LIGO}
\acrodef{AdV}{Advanced Virgo}

\newcommand{\PN}[0]{\ac{PN}\xspace}
\newcommand{\GW}[0]{\ac{GW}\xspace}
\newcommand{\SNR}[0]{\ac{SNR}\xspace}
\newcommand{\aLIGO}[0]{\ac{aLIGO}\xspace}
\newcommand{\AdV}[0]{\ac{AdV}\xspace}


\section{Introduction} 
\label{sec:introduction}

The \aLIGO~\cite{Advanced-LIGO2015} and \AdV~\cite{Advanced-VIRGO2015} detectors
carry 
the potential to observe hundreds of black-hole-binary systems per year by the time they reach design sensitivity 
(2019-20)~\cite{Abadie:2010cf,Aasi:2013wya,Dominik:2014yma}. Each binary is
characterised by the black-hole masses and spins, which we hope to measure
from \GW observations. However, we expect partial degeneracies in the dependence of the waveform on these 
parameters to limit the accuracy of their measurement~\cite{Cutler:1994ys,Poisson:1995ef,Baird:2012cu}. 

We can learn about the constituents of the binary system
through the phasing of the binary as the two objects orbit and slowly spiral toward each other. The phasing is a function of
the black-hole masses and the individual spin vectors. For a system where the spins are aligned with the binary's orbital
angular momentum we may parametrize the binary by the masses, $m_1$ and $m_2$, and the dimensionless spin parameters, $\chi_1$ and 
$\chi_2$, where the Kerr limit imposes $|\chi_i| < 1$. During the inspiral, the
leading-order \PN 
influence of the spins arises as a weighted sum
of $\chi_1$ and $\chi_2$~\cite{Poisson:1995ef,Ajith:2011ec},
\begin{equation}
\chi = \frac{ m_1 \chi_1 + m_2 \chi_2 }{ M} - \frac{76 \eta}{226} \left( \chi_1 + \chi_2 \right),
\label{eq:RedSpin}
\end{equation}
where $\eta$ is the symmetric mass ratio. 
Eqn.~(\ref{eq:RedSpin}) implies a strong degeneracy between $\chi_1$ and $\chi_2$, so that even when $\chi$ 
can be measured accurately, it will be difficult to measure the individual spins. This degeneracy is strongest for
equal-mass systems. At large mass ratios, $\chi$ is dominated 
by the spin of the larger black hole, so in these cases we may be able to measure the large black hole's spin, but the spin 
of the smaller black hole will be poorly measured, if at all.
There is also a partial degeneracy between $\chi$ and the mass ratio of the two black holes, which limits our ability to 
measure the individual masses and $\chi$~\cite{Cutler:1994ys,Poisson:1995ef,Baird:2012cu,Ohme:2013nsa}. 

When the spins also have components lying in the orbital plane, the binary's orbital plane precesses 
and the \GW signal acquires further
structure~\cite{Apostolatos:1994mx,Kidder:1995zr}. 
There are again degeneracies in the effect on the phasing, and in this case
the mean influence on the orbital precession of the four 
in-plane spin components during the course of the inspiral can be 
combined into a single ``effective precession spin'' parameter~\cite{Schmidt:2014iyl}. Once again, these degeneracies 
suggest that we may be able to measure the spin of the binary's larger black hole, but it will be difficult to accurately 
measure both black-hole spin vectors.

The preceding discussion was restricted to the black holes' inspiral. During the merger and the final black hole's 
ringdown, the \GW amplitude and phase are parameterized by the final black
hole's mass and spin, which in turn is governed by a different
combination of the progenitor masses and spins. It is conceivable that in observations of high-mass systems, where the 
inspiral and merger-ringdown contribute comparable power to the detectable signal,
that we may be able to more tightly constrain the individual masses and spins. A preliminary study of this question was 
performed in Ref.~\cite{Purrer:2013xma}. No evidence was found that two-spin effects would be measurable in 
advanced-detector observations, but the study was limited to a small number of configurations, and at the time there
was no two-spin inspiral-merger-ringdown (IMR) model available with which to systematically estimate parameter 
measurement uncertainties. We now have in hand a two-spin model that has been proposed for aligned-spin binaries, 
SEOBNRv2~\cite{Taracchini:2013rva}, which makes it possible to address this question more thoroughly. 

Here we use a simplified single-detector Bayesian framework to systematically analyze the spin information that can be
inferred from observing binary mergers.  
Although we do not perform an 
exhaustive study, and cannot rule out the possibility that particularly favourable configurations do exist, our 
basic conclusion is that only one spin parameter can be constrained by \GW observations, and 
single-spin models are sufficient for the needs of \GW astronomy during the
Advanced
detector era. 

Our study is limited to aligned-spin systems, because these are the only configurations for which a two-spin
IMR model exists that is fast enough to make parameter estimation studies feasible. Nonetheless, 
we perform some preliminary studies of precessing systems (in the inspiral
regime), and make some comments in Sec.~\ref{sec:discussion} on how well we expect our results to carry over to generic
binaries. 

We first present our methodology in Sec.~\ref{sec:methodology}, which summarises the Bayesian techniques we use to 
measure the binary parameters, the SEOBNRv2 waveform model, and our Markov-Chain-Monte-Carlo (MCMC) 
code. We then move on to our results in Sec.~\ref{sec:results}, and discuss our conclusions in Sec.~\ref{sec:discussion}.

\section{Methodology} 
\label{sec:methodology}

\subsection{Posterior probability}

In order to assess our ability to measure the source parameters $\theta$ from an
observation, we perform the following study. We simulate a \GW signal from a
binary with parameters $\theta_0$ and confront it with a variety of
signals exploring the full parameter space of possible sources. Assuming
stationary Gaussian noise of the instrument, we are
interested in the expected, noise-averaged parameter recovery. Hence we set
the noise realization to zero (i.e., our data only consist of the \GW signal
$h(\theta_0)$) while we use the instrument's noise spectral density $S_n$ in the
inner product between two signals,
\begin{equation}
 \inner{h(\theta)}{h(\theta_0)} = 4 \Re \int_{f_{\rm min}}^\infty
\frac{\tilde h(f;\theta) \, \tilde h^\ast(f;\theta_0)}{S_n(f)} \; df ~.
\label{eq:innerprod}
\end{equation}
Here, $\tilde h$ denotes the \GW signal in the Fourier domain,
$^\ast$~is the
complex conjugation, and we use the \aLIGO
``Zero-detuned High Power'' design sensitivity~\cite{T0900288} with $f_{\rm min} =
10$ Hz throughout this paper.

Given the data $d \equiv h(\theta_0)$, the likelihood ratio between the
hypothesis that a signal with parameters $\theta$ is contained in $d$, or $d$
is pure Gaussian noise, is \cite{Finn:1992xs}
\begin{equation}
 \Lambda =
\frac{\exp(-\inner{h(\theta_0-h(\theta)}{h(\theta_0)-h(\theta)}/2)}{\exp(-\inner
{ h(\theta_0) } { h(\theta_0) } /2) }. \label{eq:likelihood}
\end{equation}

We simplify Eqn.~(\ref{eq:likelihood}) by expanding the linear inner
product and maximizing the likelihood over the template norm $\rho =
\sqrt{\inner{h(\theta)}{h(\theta)}}$, 
\begin{equation}
\hat \Lambda = \max_\rho \Lambda = \exp\left( \frac{\rho_0^2}{2}
\inner{\hat h(\theta)}{\hat h(\theta_0)}^2 \right) ~. \label{eq:maxLambda}
\end{equation}
Here, $\rho_0 = \sqrt{\inner{h(\theta_0)}{h(\theta_0)}}$ is the simulated \SNR,
and 
\begin{equation}
 \hat h = \frac{h}{\sqrt{\inner{h}{h}}}
\end{equation}
denotes the normalized waveform.  

Bayes' theorem allows us to express the posterior probability of the
source parameters, given the data, as the product of the likelihood $\hat
\Lambda(\theta)$ with the prior probability $\pi(\theta)$,
\begin{equation}
 \mathcal P(\theta) = \frac{\hat \Lambda (\theta) \,
\pi(\theta)}{\mathcal P_0},
\end{equation}
where $\mathcal P_0$ can be interpreted as a simple normalisation factor such that
\begin{equation}
 \int \mathcal P(\theta) \, d\theta = 1.
\end{equation}
When we are interested in the posterior probability for individual
parameters, we present the marginalized posterior by integrating over all
parameters that we are not interested in. Formally, our results are then based
on a mixture of a likelihood that was maximized over extrinsic parameters
(distance, orientation, sky location) which all contribute to the template norm,
cf.~Eqn.~(\ref{eq:maxLambda}), and a posterior that is marginalized over intrinsic
parameters.

The intrinsic parameters we vary are the chirp mass $M_c$, the symmetric mass
ratio $\eta$, the dimensionless, aligned spin components $\chi_i$ ($i = 1,2$),
as well as a reference (or ``coalescence'') time $t_c$ and phase $\phi_c$,
\begin{eqnarray}
 \theta &=& \{ M_c, \eta, \chi_1, \chi_2, t_c, \phi_c \}, \\ \nonumber
 M_c &=& \frac{(m_1 \, m_2)^{3/5}}{(m1+m2)^{1/5}},  \\ \nonumber
 \eta &=& \frac{m_1 \, m_2}{(m_1 + m_2)^2}, \\ \nonumber
 \chi_i &=& \frac{\vec S_i \cdot \hat L}{m_i^2}.
\end{eqnarray}
Here, $m_i$ are the individual black hole masses, $\vec S_i$ are their spin
vectors and $\hat L$ is the direction (unit length) of the orbital angular
momentum. We only
consider spins that are either aligned or anti-aligned with $\hat L$. 

In order to optimize the calculation of $\mathcal P(\theta)$, we use the fact
that we can marginalize analytically over $\phi_c$. The dominant
mode of non-precessing signals obeys
\begin{equation}
  h (\phi_c + \Delta \phi) =  h (\phi_c) \; e^{-i2 \Delta \phi},
\end{equation}
which allows us to express
\begin{eqnarray}
 \frac{1}{2\pi} \int_0^{2\pi} \hat \Lambda \; d\phi_c &=&
 \frac{1}{2\pi} \int_0^{2\pi} e^{8 \rho_0^2 \cos^2 (2\phi_c)
\left \vert  \mathcal O \right
\vert^2}\; d\phi_c \\
&=&  e^{4 \rho_0^2 \vert \mathcal O \vert^2} \; I_0\left(4 \rho_0^2 \vert
\mathcal O \vert^2\right),
\end{eqnarray}
where $I_0$ is the Bessel function of the first kind and $\mathcal O$ is the
complex integral that is part of the the inner product,
\begin{equation}
 \mathcal O = \int_{f_{\rm min}}^\infty
\frac{\tilde h(f) \, \tilde h_0^\ast(f)}{S_n(f)} \; df~.
\label{eq:complex_overlap}
\end{equation}

In addition, we can efficiently marginalize over the time $t_c$ by noting that
\begin{equation}
 \tilde h(t_c + \Delta t) = \tilde h(t_c) \; e^{i 2\pi f \Delta t}.
\end{equation}
Therefore, we can calculate $\mathcal O$ for a range of time shifts by using
fast algorithms for the discrete inverse Fourier transformation. By summing the
results over time, we effectively marginalize over $t_c$ as well, which means
our codes only have to sample the physical mass and spin related parameters. A
similar strategy has been described by W.~Farr in an internal technical
document \cite{Farr:T1400460}.

We note that \emph{maximizing} over time and phase is computationally even
faster, and our code is able to perform either the maximization or
marginalization over time and phase efficiently as a result of the above
calculations. Since we found virtually indistinguishable results with both
methods, we used time- and phase-maximized posteriors in the majority of
cases we report here.

\subsection{MCMC code} 
\label{sub:mcmc_code}

We use the ensemble sampler \emph{emcee}~\cite{ForemanMackey:2012ig} together
with the LALSuite~\cite{LAL-web} library that contains the implementation of the
waveform models discussed in Sec.~\ref{sub:reduced_order_model_for_seobnrv2}.

The simulations use the single-detector likelihood described in
Eq.~(\ref{eq:maxLambda}). This is based on the inner product
(\ref{eq:innerprod}) which we calculate as a discrete inverse Fourier
transformation (see the discussion in the previous section). We discretize
the integral with a frequency spacing
of $\Delta f = 0.5 \text{Hz}$, which is larger than the inverse duration of
some of the signals we consider. However, we only need to resolve the
\emph{correlations} between
$h(\theta)$ and $h(\theta_0)$, and in numerical tests these turned out to have
significant support only over a small range of time shifts if the optimally
aligned waveforms agree well, which is the region we are interested in
here. In order to increase
the time-domain resolution for maximization or marginalization over $t_c$, we
use zero-padding of the signal and template by a factor of two. 

In general we use $\sim$10000 iterations with $100$ MCMC chains. The first
half of the samples is discarded to remove the burn-in of the chains which are
not draws from the posterior distribution. The runs are further checked by
examining the homogeneity of the trace of all chains.

As in Ref.~\cite{Veitch:2014wba} we choose uniform priors in the component
masses with a range of $1 M_\odot \leq m_i \leq 300 M_\odot$. We sample only
half of the mass space and assume $m_2 \geq m_1$ and add a cutoff in total mass
$M_\text{tot} \leq 500 M_\odot$. The priors on the aligned spins are taken to be
uniform in $\chi_i \in [-1, 0.99]$.

In addition, two simulations with the precessing \PN SpinTaylorT4 model as
implemented in LALSuite~\cite{LAL-web} were performed with the nested sampling
code from \texttt{lalinference}~\cite{Veitch:2014wba} for a three-detector 
\aLIGO-\AdV setup.
We used a sampling rate of $4096 \text{Hz}$, no amplitude corrections, zero
noise, the \aLIGO~\cite{T0900288} and \AdV~\cite{2012arXiv1202.4031M} design
PSDs with a lower frequency cutoff of $40 \text{Hz}$ and a network \SNR of 30.


\subsection{Reduced order model for SEOBNRv2} 
\label{sub:reduced_order_model_for_seobnrv2}

To explore the measurability of aligned component spins for complete IMR waveforms over a large parameter space region we use a \emph{reduced order model} (ROM)~\cite{Purrer:2015tud} ``SEOBNRv2\_ROM'' of the spin-aligned effective-one-body model ``SEOBNRv2''~\cite{Taracchini:2013rva}. At the moment SEOBNRv2 is the only two-spin IMR model available, but due to its high computational cost it cannot be used directly for parameter estimation studies that routinely require millions of likelihood evaluations. This study has only become feasible with the availability of the ROM~\cite{Purrer:2015tud} which is constructed using extensions to the techniques described in Ref.~\cite{Purrer:2014fza}. This ROM provides speedups on the order of several thousands over SEOBNRv2. 

The SEOBNRv2 model was found to be accurate to a mismatch of $1\%$ against 38 NR waveforms from the SXS 
collaboration~\cite{SXS:catalog}, for total masses $M \in [20, 200] M_\odot$ and the \aLIGO design power spectral density (PSD)~\cite{T0900288}. Recent studies have shown that while SEOBNRv2 is extremely accurate, it disagrees with new BAM~\cite{Husa:2015iqa} and SXS~\cite{SXS-Chu-catalog} NR waveforms at high aligned spin~\cite{Khan:2015jqa,Prayush-Chu-SEOBNR-Phenom-comparison}.

The ROM~\cite{Purrer:2015tud} has a worst mismatch against SEOBNRv2 of $\sim$1\%, but in general mismatches are better than $\sim$0.1\%. It covers the entire SEOBNRv2 parameter space $0.01 \leq \eta \leq 0.25$ and $-1 \leq \chi_i \leq 0.99$ for compact binaries of total mass $M_\mathrm{tot} \geq 2 M_\odot$ and the full \aLIGO design sensitivity starting at $10$ Hz.

We note that the LAL implementation~\cite{LAL-web} of SEOBNRv2 and SEOBNRv2\_ROM
only provide the $l=m=2$ mode of the \GW signal. Currently, no models are
available that include both spin and higher modes. Since we also use
SEOBNRv2\_ROM for the target signals we cannot investigate the effects of higher
modes in this study.

The waveform used here also does not include effects of precession of the
orbital plane and of the black-hole spins. Precession is driven by mainly a
single measurable parameter~\cite{Schmidt:2014iyl,Hannam:2013oca} that describes
spin in the orbital plane. At total masses where the merger ringdown contributes
significantly to the overall power the number of precession cycles is very
small. Therefore we do not expect precession to change the qualitative picture
at those total masses, as we will discuss in Sec.~\ref{sec:discussion}.




\section{Results} 
\label{sec:results}

We now use the methodology described in Sec.~\ref{sec:methodology} to study the accuracy with which we can
measure the individual spins of aligned-spin binaries. We focus on total masses where both the inspiral and merger-ringdown 
contribute a significant amount of power to the \GW signal in \aLIGO. The
nominal mass we choose is $50 M_\odot$.
If we make a crude split between inspiral and merger-ringdown at the Schwarzschild ISCO frequency, then at this mass
a nonspinning equal-mass binary produces 60\% of its detectable power during the inspiral, and 40\% during the 
merger and ringdown.
We use the \aLIGO ``Zero-detuned High Power'' design sensitivity~\cite{T0900288} with a lower frequency cutoff of $f_{\rm min} = 10$ Hz.

We consider mass ratios of $q=1,4$ and 10, and a range of (equal) aligned spins $\chi_i = -0.9, 0, 0.5, 0.9$. 
We explore parameter recovery with two models: the double aligned-spin SEOBNRv2 model
 and a single-spin SEOBNRv2 model,
i.e. a model that assumes equal-spin waveforms 
$\chi_1 = \chi_2$, which can be expressed in terms of the reduced spin parameter $\chi$.
Instead of $\chi$ we present results in terms of a \emph{rescaled reduced spin}~\cite{Khan:2015jqa}
\begin{equation}
  \label{eq:chi_hat}
  \hat \chi := \frac{\chi}{1 - 76\eta/113},
\end{equation}
which takes values in $[-1,1]$.
We choose an \SNR of 30, which is considered optimistic but not unreasonable for
\aLIGO observations.
Note that if we assume a uniform volume distribution of sources, and a threshold
\SNR of $\sim$10 for detection, then since
the \SNR is proportional to the source distance, only $(10/30)^3$ of signals
will be close enough to have an \SNR higher than
30, i.e., $\sim$96\% of signals will have \acp{SNR} below 30. Similarly, only
one in a hundred observations will have an \SNR greater 
than 50, and one in a thousand greater than 100.

Once we have presented results for our nominal choices (equal spins, $50\,M_\odot$, and an SNR of 30), we consider
the effect on our results of varying each of these in turn; we look at unequal spins in Sec.~\ref{sub:unequal_spins},
the effect of varying the total mass in Sec.~\ref{sub:dependence_on_total_mass}, and the dependence on SNR in 
Sec.~\ref{sub:dependence_on_snr}, to determine at which SNR we may be able to constrain the spin on the smaller
black hole.

\subsection{Equal-spin binaries}
\label{sub:equal_spin}

\begin{figure*}[htbp]
	\centering
		\includegraphics[width=0.45\textwidth]{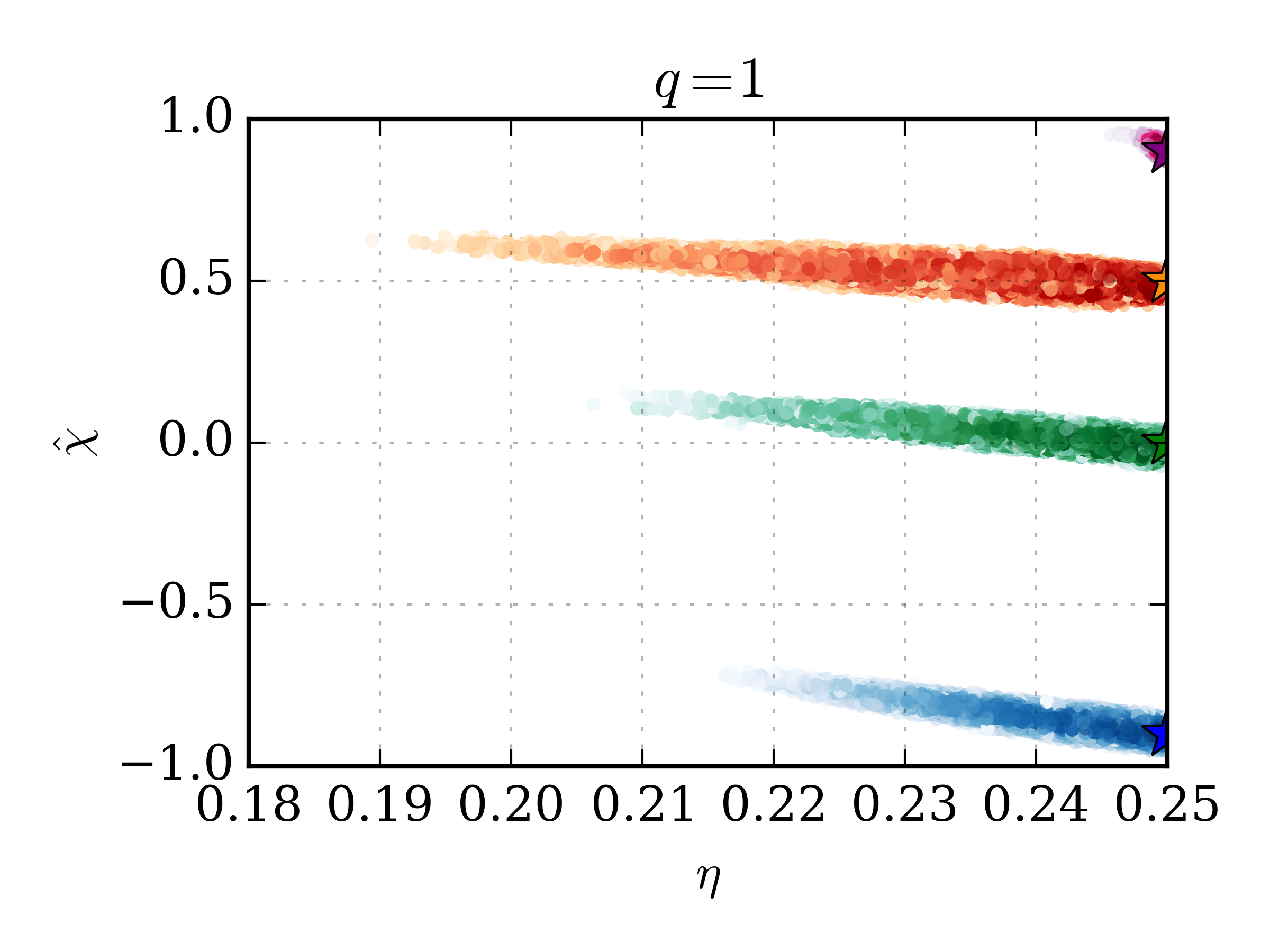}
		\includegraphics[width=0.45\textwidth]{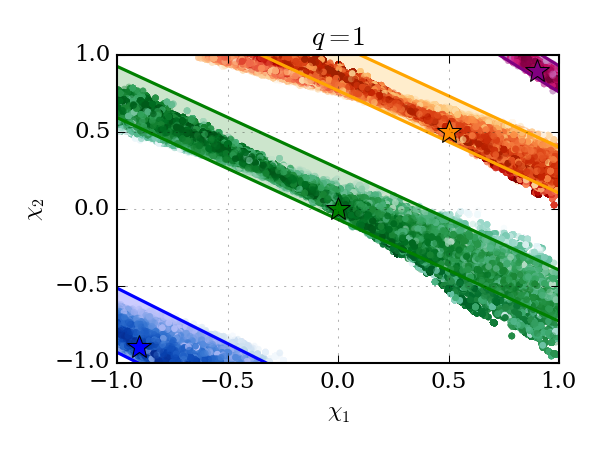}
		\includegraphics[width=0.45\textwidth]{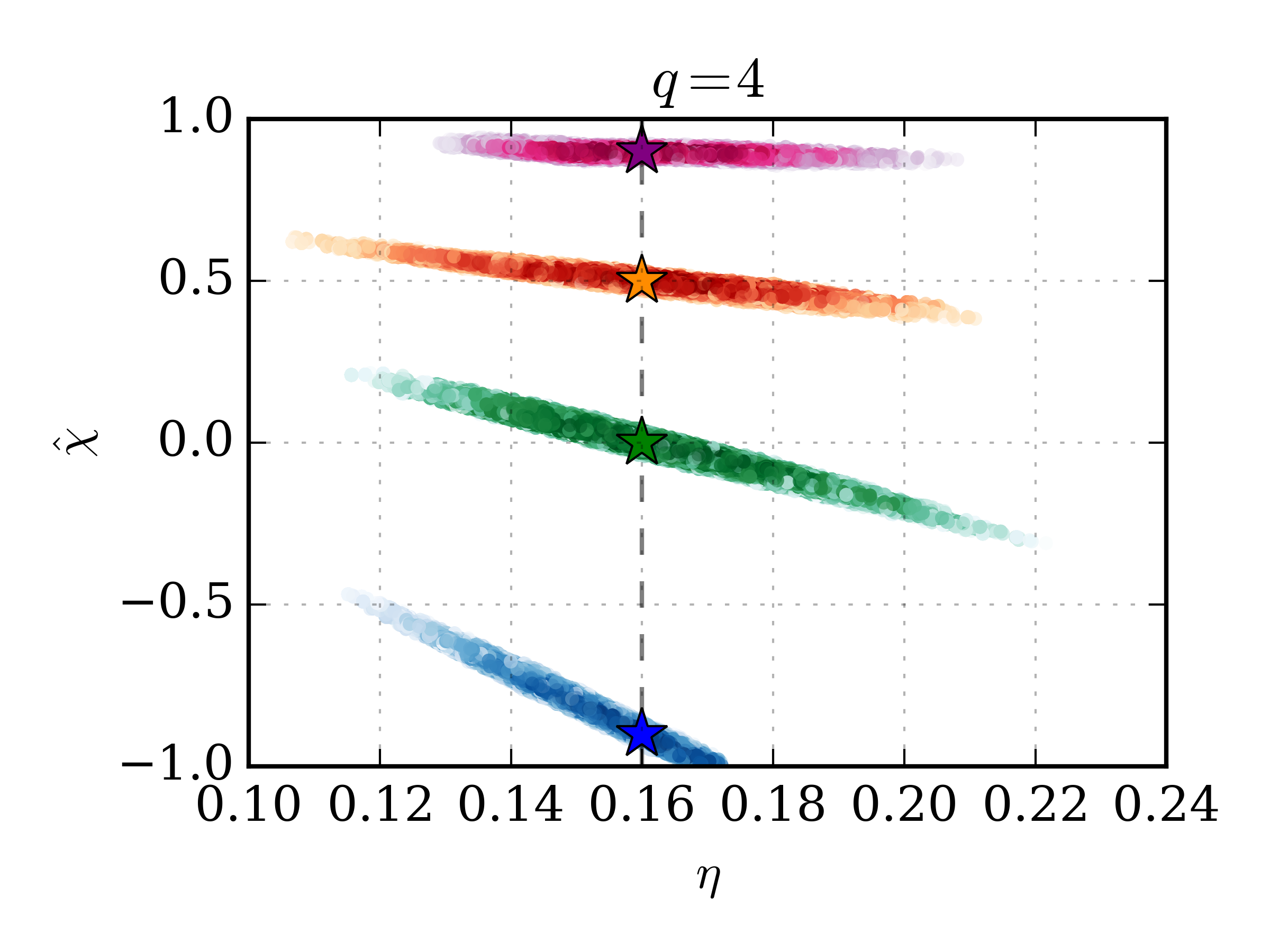}
		\includegraphics[width=0.45\textwidth]{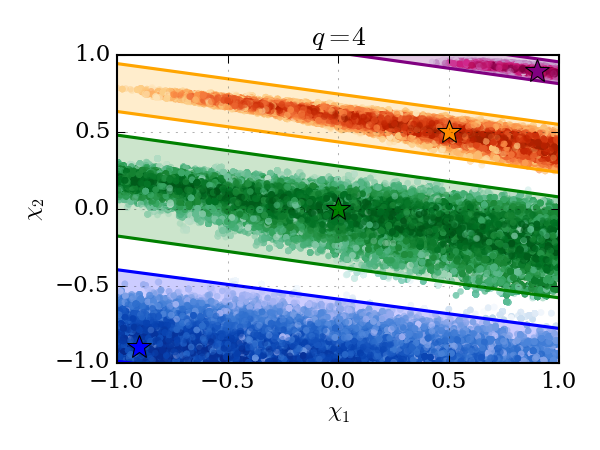}
		\includegraphics[width=0.45\textwidth]{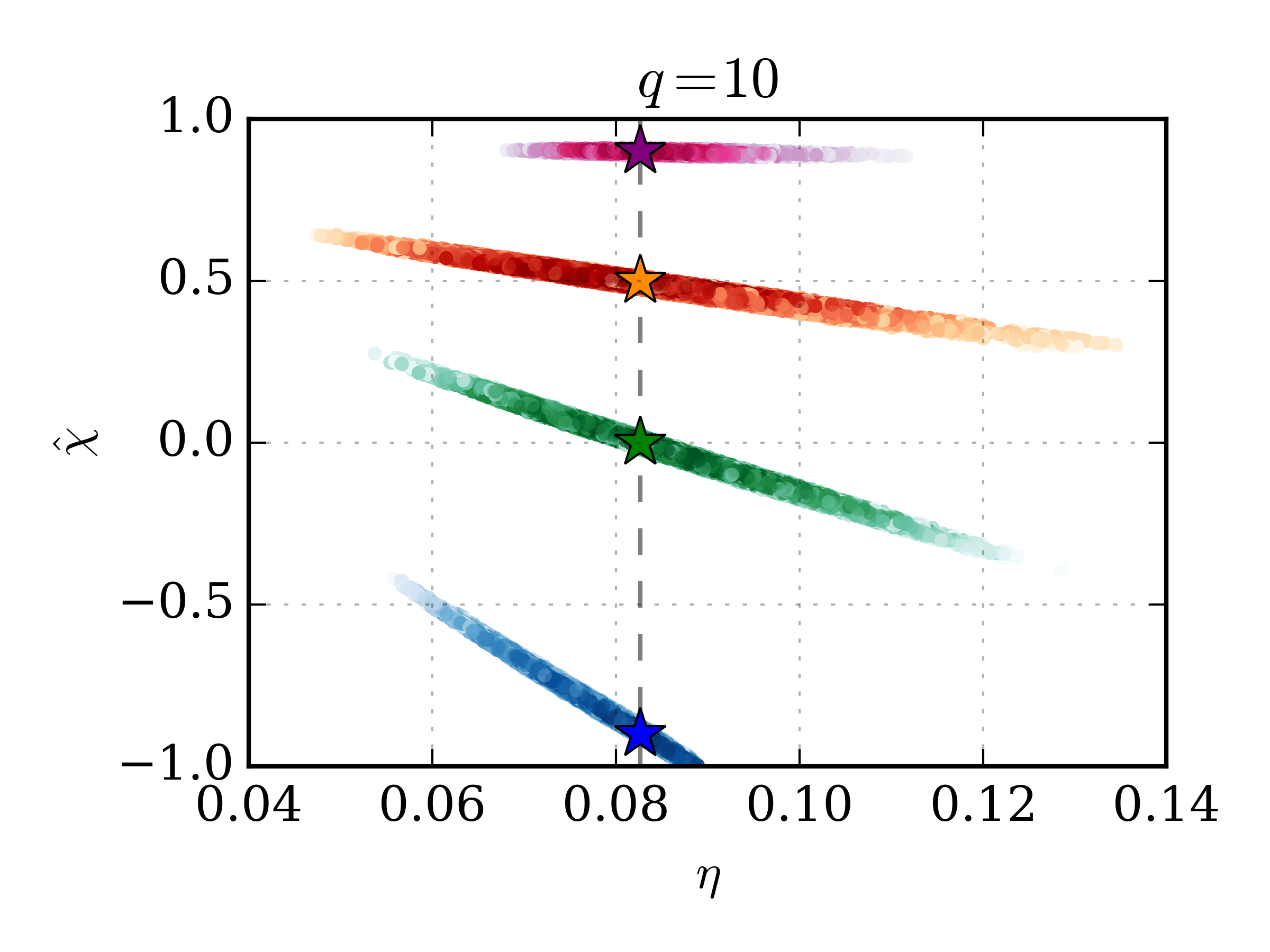}
		\includegraphics[width=0.45\textwidth]{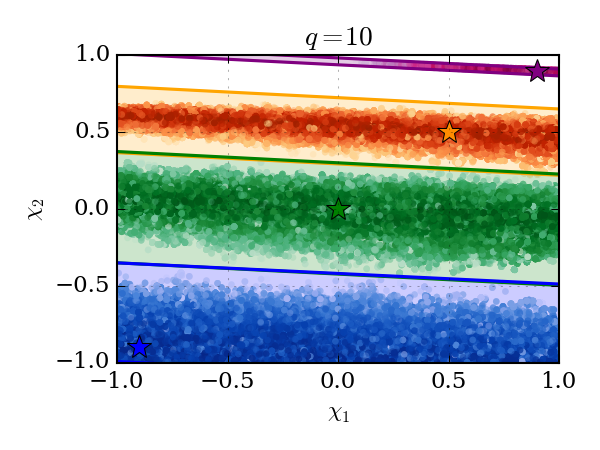}
\caption{Symmetric mass-ratio $\eta$ vs rescaled reduced spin $\hat\chi$ (left), and component spin $\chi_1$ vs $\chi_2$ (right) posteriors for configurations with mass ratios $q=1,4,10$ (top to bottom) at total mass $50 M_\odot$ and SNR 30. Each panel shows configurations with equal aligned spins $\chi_1=\chi_2 = -0.9, 0, 0.5, 0.9$ (blue, green, orange, magenta) with the true parameters indicated by a star symbol.
In the right column we show in addition to the $\chi_1$--$\chi_2$ samples colored bands that are delineated by two lines of constant reduced spin. These are obtained from the posteriors of the single-spin model.
}
	\label{fig:eta_chi_chi1_chi2_posteriors_Mtot50Msun}
\end{figure*}

Our key results for $50\,M_\odot$ binaries are shown in
Fig.~\ref{fig:eta_chi_chi1_chi2_posteriors_Mtot50Msun}.
The left column of panels show the posteriors for the reduced spin with respect to the symmetric mass ratio, while
the right panel shows the posteriors for $\chi_2$ versus $\chi_1$. The plots show all of the samples from the posterior
probability distribution function so that the full structure of the marginal PDF can be seen. When making comparisons between configurations we show $95\%$ credible regions $\mathcal{C}$ which are a subset of the parameter space that includes $95\%$ of the posterior probability, $\int_\mathcal{C} \mathcal{P}(\theta) d\theta = 0.95$. In most cases the extent of 95\% credible regions can be estimated from the full set of samples by eye.

Much of what we can conclude about our ability to measure the individual black-hole spins can be 
inferred from this figure; the remainder of this paper will be devoted to demonstrating that the general trends
we observe here are generic. Our observations are as follows:

Consider first the panels on the left, which show the reduced spin with respect to mass ratio. The posterior samples
are arranged in strips. This is consistent with the degeneracy between mass ratio and spin that has been 
discussed in detail in previous work~\cite{Baird:2012cu,Ohme:2013nsa,Poisson:1995ef,Hannam:2013uu}. During the inspiral
the approximate degeneracy is between the mass ratio and the reduced spin (see the discussion in 
Ref.~\cite{Baird:2012cu}), while during the ringdown the degeneracy is between systems with the same final mass and
spin. For the systems shown in Fig.~\ref{fig:eta_chi_chi1_chi2_posteriors_Mtot50Msun}, both degeneracies play a
role. We will see in Sec.~\ref{sub:dependence_on_total_mass} how these uncertainty regions vary with respect to total mass.

The tilt of the $\eta$-$\hat\chi$ posteriors depends on the spin and on the mass ratio. The regions are almost parallel to 
the $\eta$ axis for high aligned spins and have the largest slope for high anti-aligned spins.  
The variation in the tilt of the posteriors with respect to spin becomes more pronounced as we move to 
higher mass ratios. 
Thus, we can measure the reduced spin best for configurations with high aligned spin or equal-mass systems, 
and worst if the spin is highly anti-aligned and the system at higher mass ratio. We note that this effect is due to
the parameter dependence of the signals during the inspiral; we will see in Sec.~\ref{sub:dependence_on_total_mass} that as we move to 
higher-mass systems and more of the signal is from the merger and ringdown, that these regions rotate.

Now consider the right-hand panels, which show the same results, but with $\chi_1$ plotted against $\chi_2$. 
The solid lines indicate the results from using a single-spin model
parameterized with respect to the effective spin; these lines represent constant values of the
reduced spin, and their orientation arises purely from the definition Eq.~(\ref{eq:RedSpin}). 
We now see that the individual spins are very poorly constrained, and the posterior samples extend over the 
full spin range possible for each value of the reduced spin. 

For equal-mass systems, lines of constant reduced
spin are diagonal, and we measure each spin equally poorly. The spin measurements are constrained only by 
the Kerr limit. This means that we can only measure them accurately if they are both near-extremal and with the 
same orientation; this is clear from the configurations with spins $\chi_1 = \chi_2 = \pm 0.9$. At the other extreme,
if the reduced spin is small, as in the case where $\chi_1 = \chi_2 = 0$, then the measurement is consistent with 
\emph{any} magnitude of $\chi_1$ or $\chi_2$, with only the requirement that the other spin be of similar magnitude
and in the opposite direction. 

As the mass ratio increases, the spin of the larger black hole increasingly
dominates the \GW phasing, and this
causes the lines of constant reduced spin to rotate. At high mass ratios, the spin of the larger black hole becomes a
better approximation of the value of the reduced spin. At mass ratio 1:10, we see that we are now able to better 
measure the spin of the larger black hole; its uncertainty has been reduced to roughly a factor of two of the 
uncertainty in the reduced spin. The spin of the smaller black hole remains unconstrained, and in fact it is poorly 
constrained even for systems with a large reduced spin. 

This tells us that our best hope of accurately measuring the spin of both black holes in a binary is for an equal-mass
system with both spins near-extremal and both aligned (or both oppositely aligned) to the orbital angular momentum.
For high mass-ratio systems, or those with a small-to-moderate value of the reduced spin, the small black hole's
spin is difficult to measure at all. However, for large mass-ratio systems, we \emph{can} measure the large black 
hole's spin, regardless of its value. 

Based on the results from the single-spin model that are overlaid on the right-hand panels of 
Fig.~\ref{fig:eta_chi_chi1_chi2_posteriors_Mtot50Msun}, we may reasonably ask: what does a two-spin model
tell us that we do not already learn from a simpler single-spin model, together with the constraints that follow from the
definition of the reduced spin? The mass-ratio 1:4 results with moderate
spins suggest that we will obtain a slightly stronger bound on the spin of the
smaller black hole with a two-spin model, 
but the improvement is not dramatic. We will return to this point in subsequent sections, where we consider
systems with unequal spins, varying masses, and higher values of the \SNR.

\subsection{Unequal spins} 
\label{sub:unequal_spins}

\begin{figure*}
  \centering
 
\includegraphics[width=.45\textwidth]
{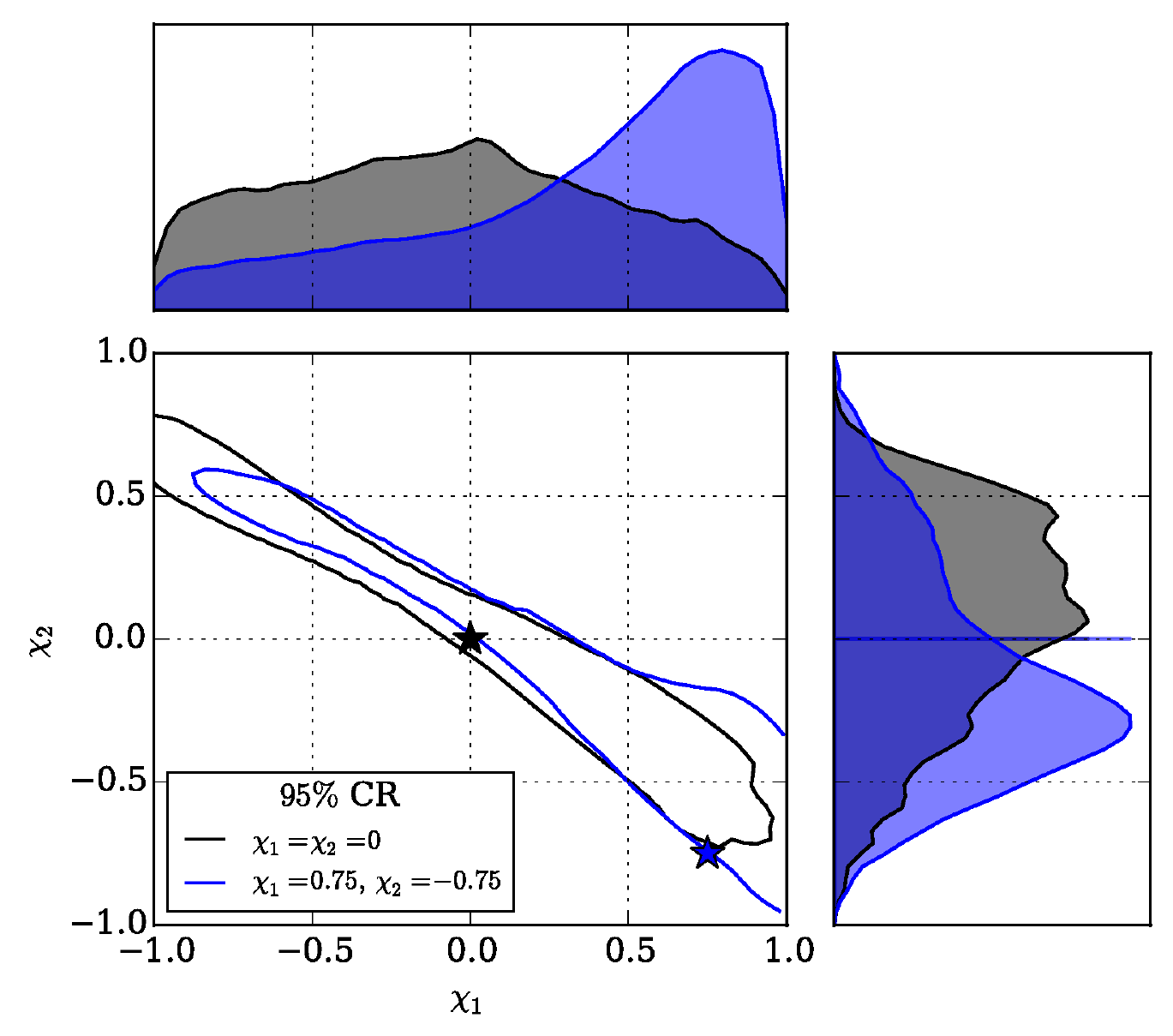}
	\includegraphics[width=.45\textwidth]{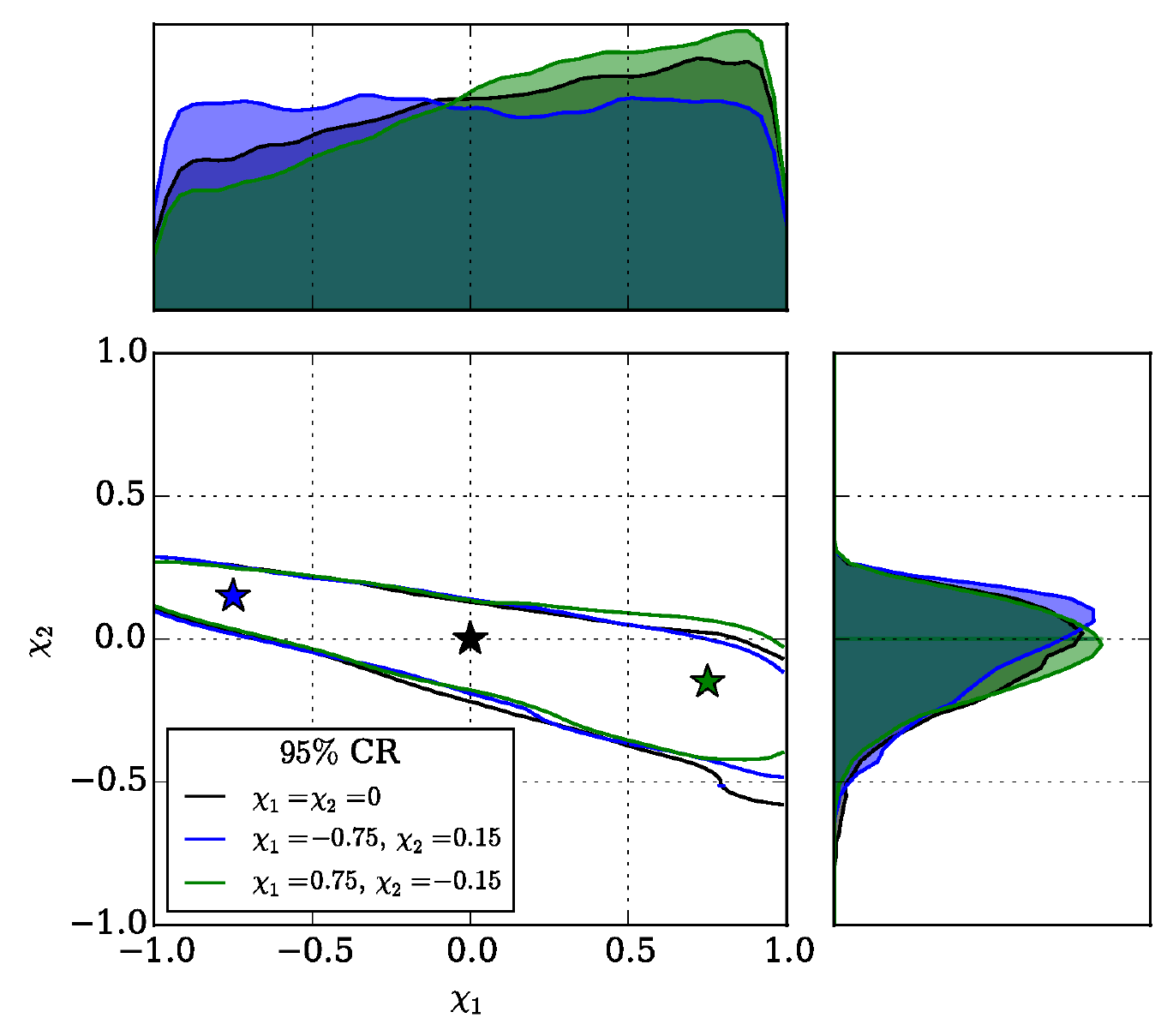}
  \caption{$95 \%$ credible regions for unequal spins, but identical reduced
spin $\chi$. The configurations have \SNR $30$, and $M_\text{tot} = 50M_\odot$
and mass-ratio $q=1$ (left) and $q=4$ (right). While the signal is symmetric
under exchange of the masses for $q=1$, the prior is not.
}
  \label{fig:unequal_spin_3_KDE_q4_50Msun}
\end{figure*}

So far we have restricted ourselves to configurations with equal aligned spins $\chi_1 = \chi_2$. We have found
that the reduced spin $\chi$ is measured well, but the individual spins are not, beyond constraints implied by the
Kerr limit. In this section we illustrate that the situation does not appear to change even if the spins are unequal.

As our example we consider three configurations at mass-ratio 1:4, all with the same reduced spin $\chi = 0$. These
systems have a total mass of 50\,$M_\odot$. 
In one configuration both black holes have no spin, and in the others $\chi_1 = \pm 0.75$, with $\chi_2$ chosen such
that the reduced spin is still zero. 

Fig.~\ref{fig:unequal_spin_3_KDE_q4_50Msun} shows the two-dimensional $95 \%$ credible regions for the 
component spins and the one-dimensional probability distributions of the spins. 
It is evident that the credible regions are almost identical. 
We first look at results for mass-ratio $q=4$ (right panel). While the PDFs of the spins peak
close to the true values, they all have almost the same support. In the presence of detector
noise these differences are unlikely to be measurable.
At mass-ratio $q=1$ (left panel) the credible regions cover a similar region with some difference near the corners of the spin plane, however, there is a marked difference in where the 1D PDFs of the spins peak depending on the spins of the signal. These details are not available from a single-spin model where the distribution for $\chi$ yields a band that is close to parallel to the credible regions.
In contrast, a two-spin model allows one to compute the maximum a posteriori
probability (MAP) for the spin components which can be a useful point estimate
of the PDFs at least in theoretical studies or sufficiently high \SNR. With the
inclusion of instrumental noise the MAP is probably less useful because it very
likely is not located at the peak of the zero-noise posterior. 
For sufficiently high \SNR the extent of the spin posterior from a two-spin
model
can be shorter than the band obtained from the single-spin model which always
covers the full range $[-1,1]$ in the spin of the small black hole and so, the
single-spin model would overestimate the measurement uncertainty in $\chi_1$.

\subsection{Dependence on total mass} 
\label{sub:dependence_on_total_mass}

\begin{figure*}[htbp]
	\centering
		\includegraphics[width=0.3\textwidth]{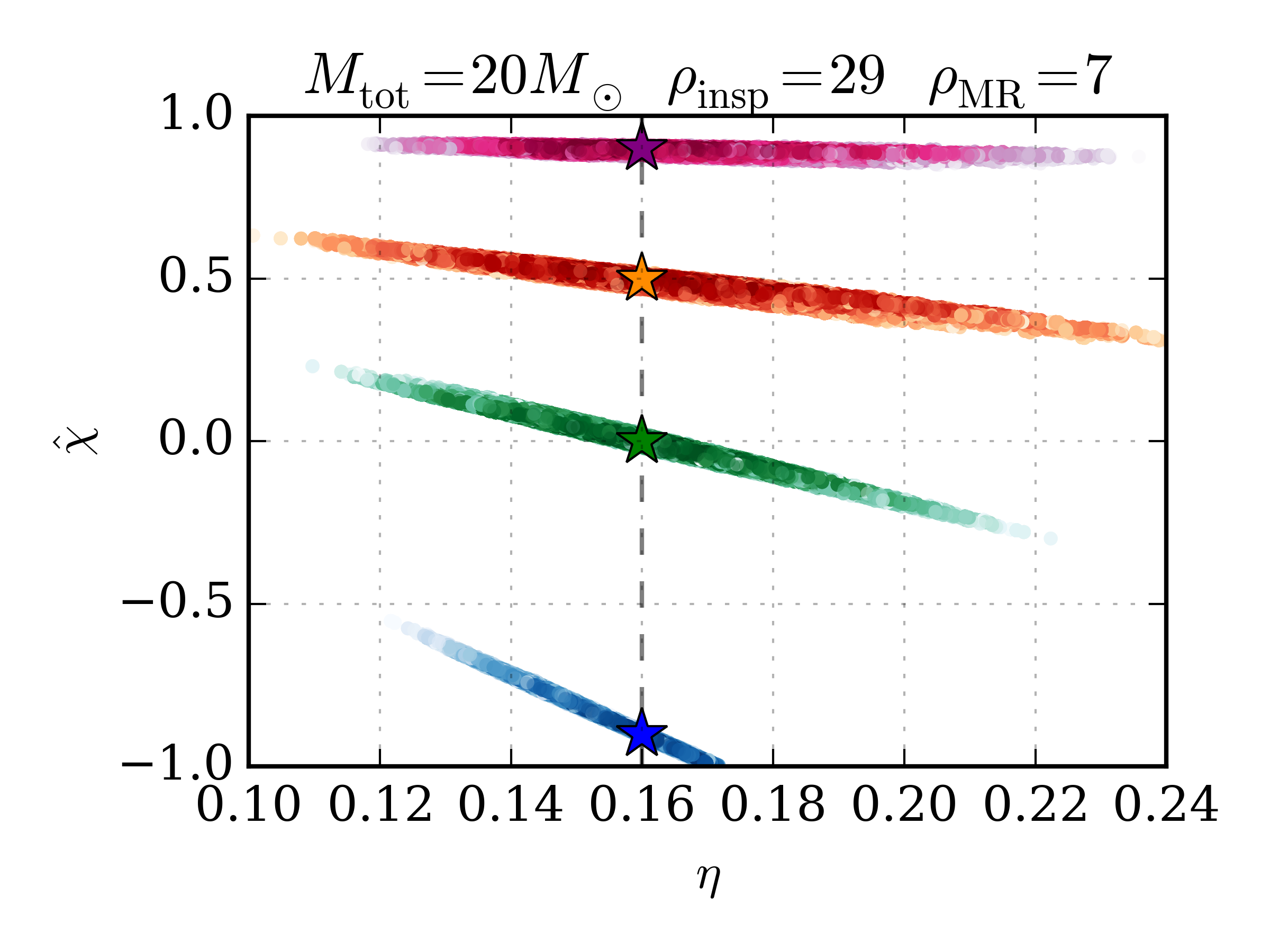}
		\includegraphics[width=0.3\textwidth]{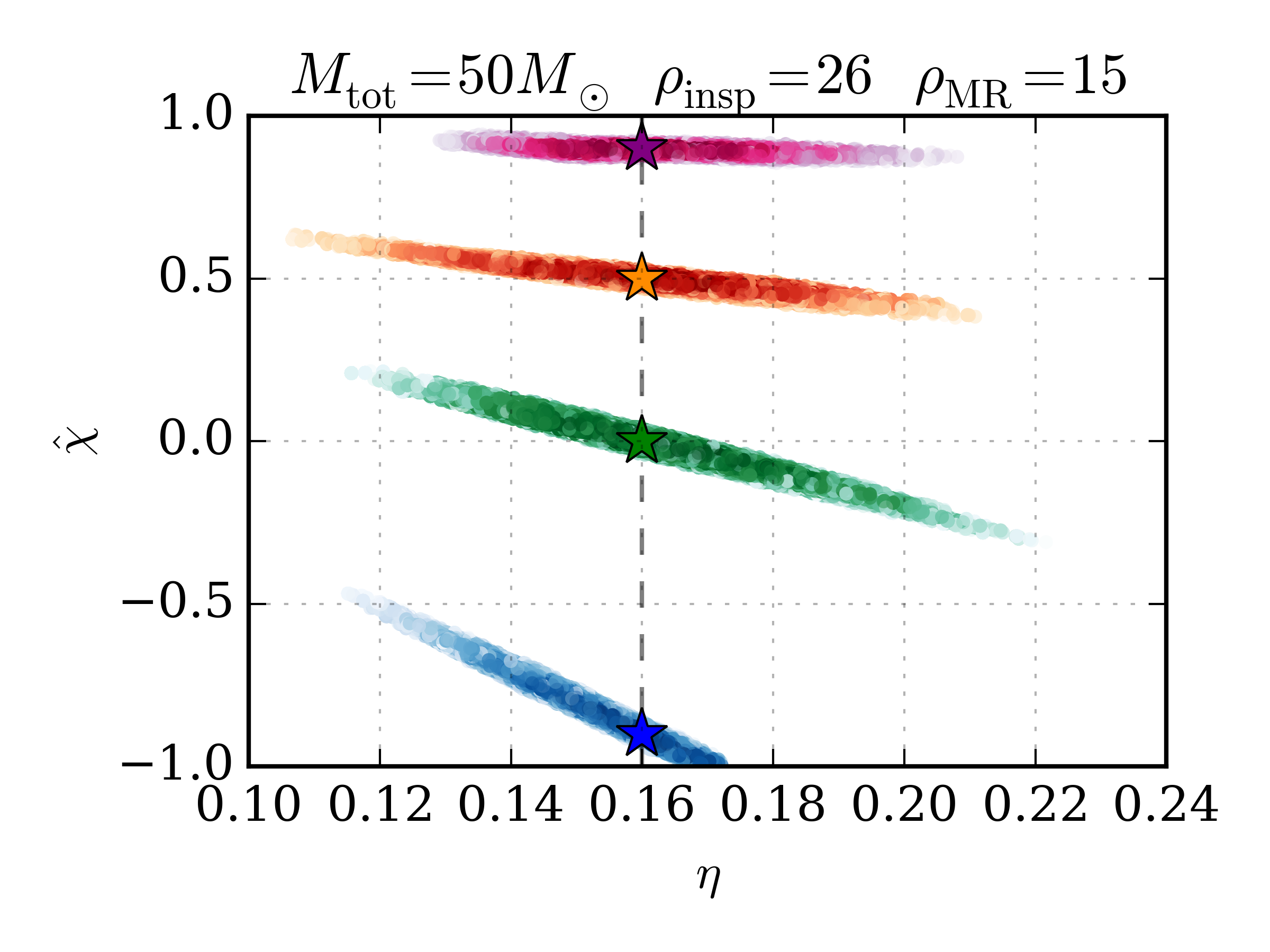}
		\includegraphics[width=0.3\textwidth]{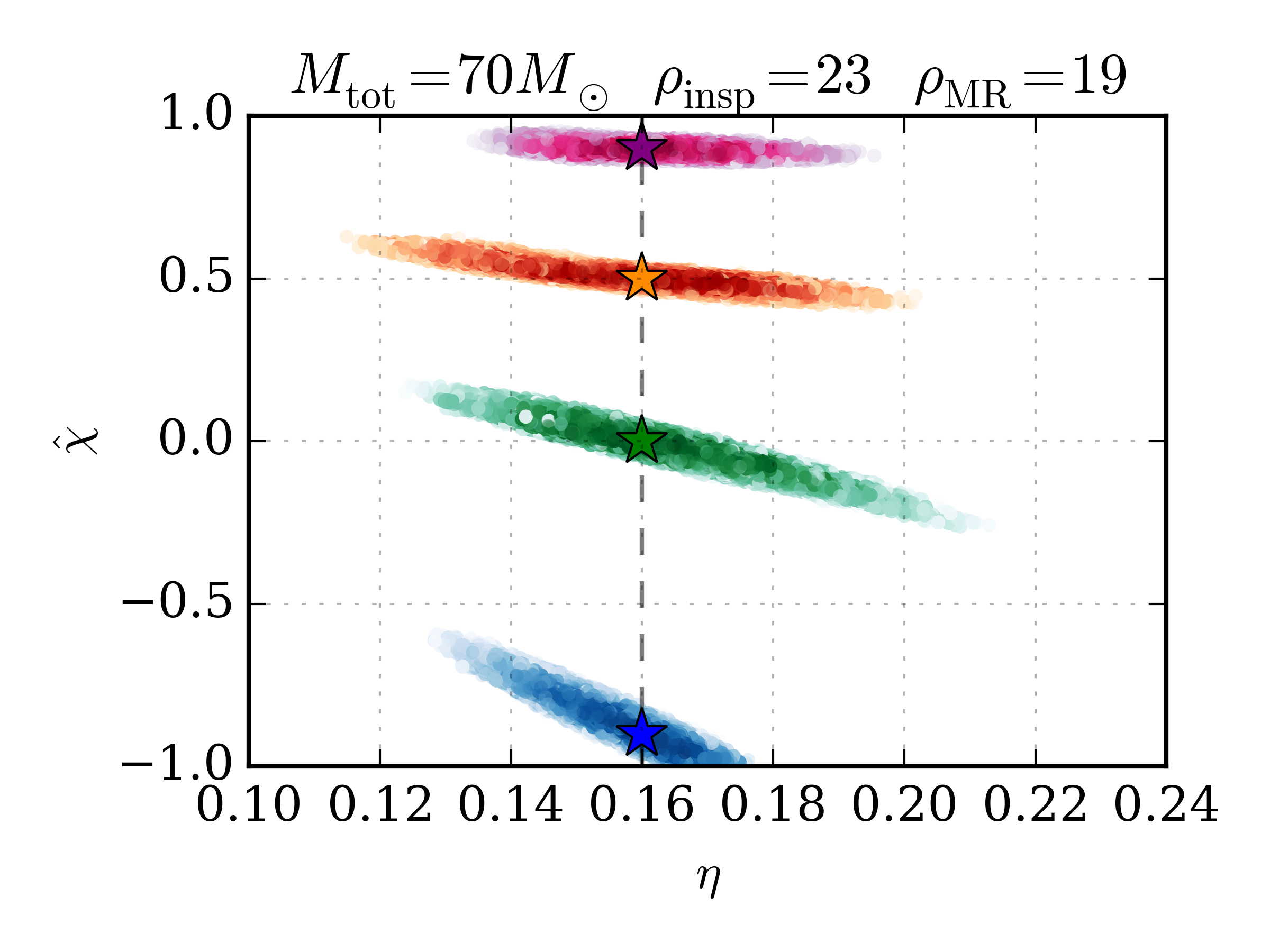}
		\includegraphics[width=0.3\textwidth]{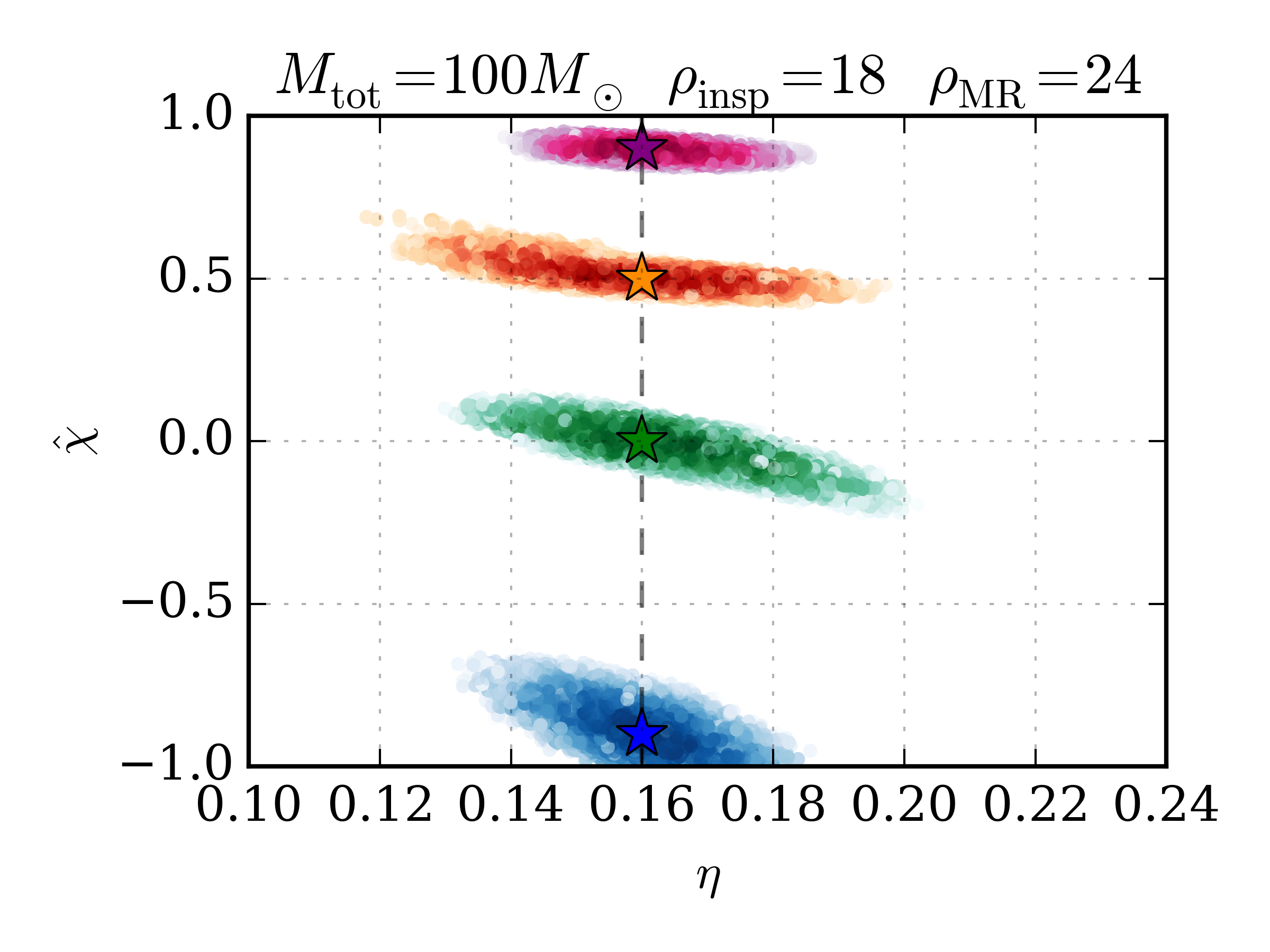}
		\includegraphics[width=0.3\textwidth]{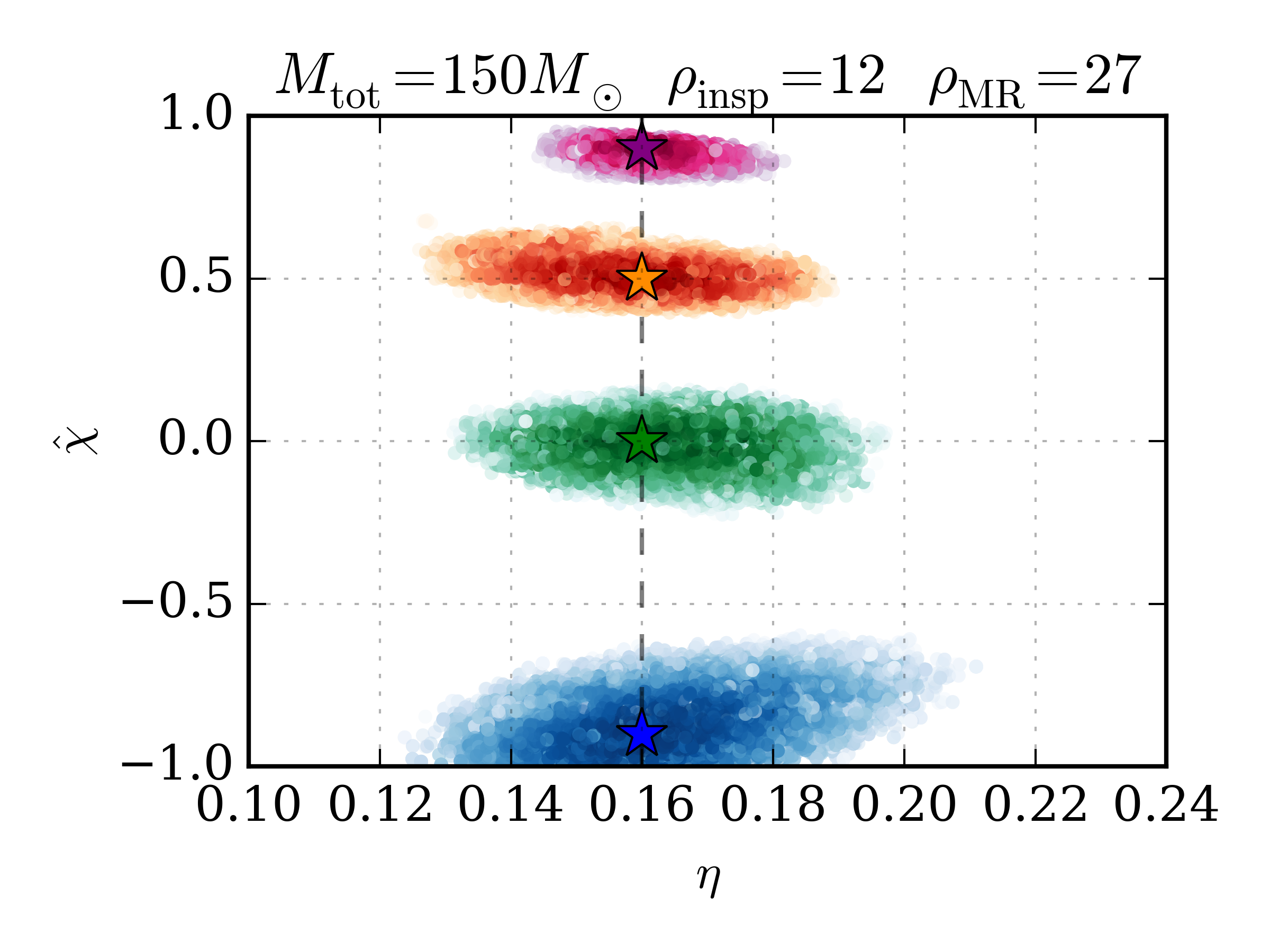}
		\includegraphics[width=0.3\textwidth]{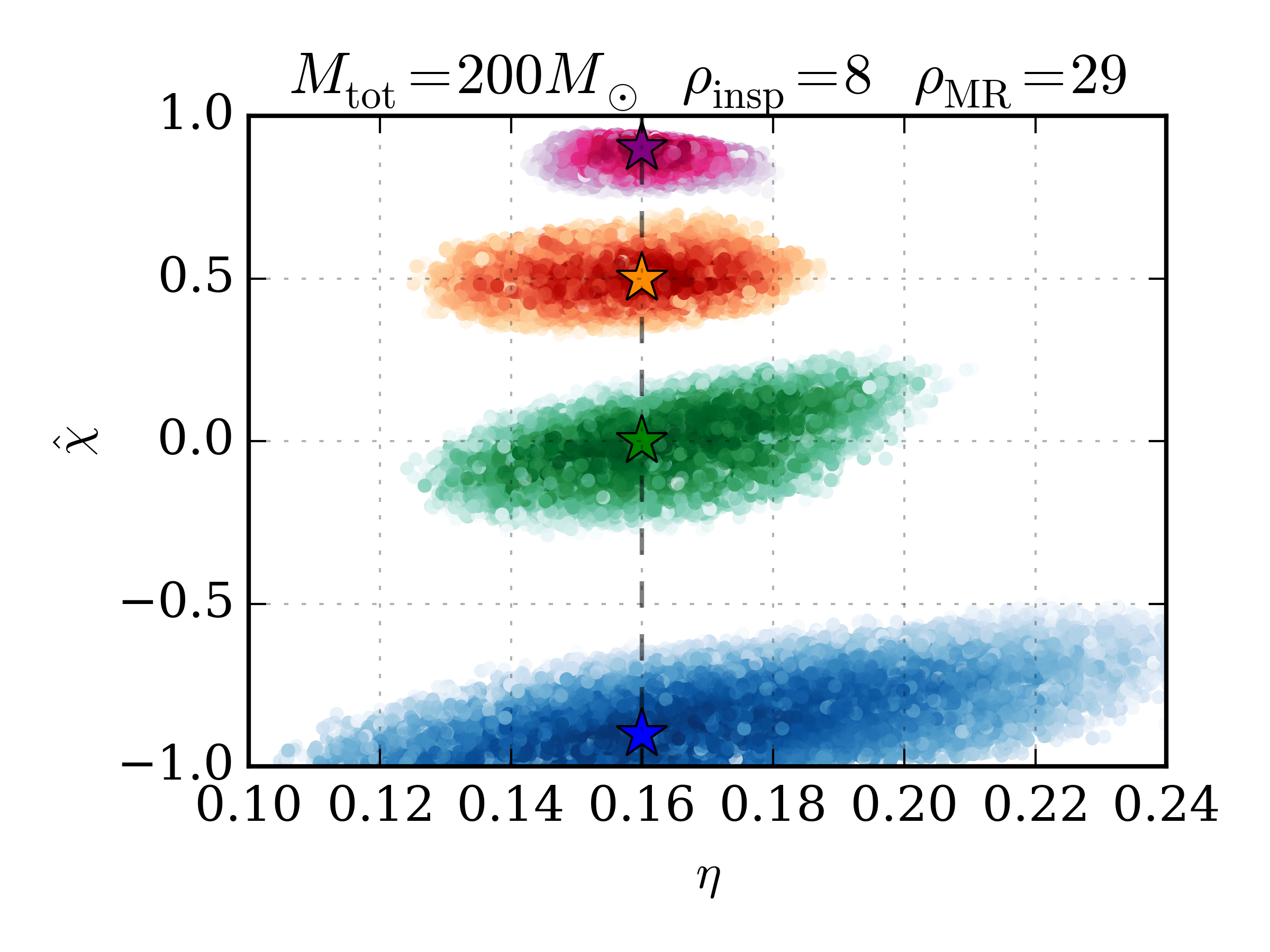}	
	\caption{$\eta$-$\hat\chi$ posteriors for mass-ratio $q=4$ and a range
of total masses. We give \acp{SNR} in the inspiral (up to the Schwarzschild
ISCO) and merger-ringdown (from the ISCO onwards) for nonspinning
configurations. The total \SNR is 30 in all cases.}
	\label{fig:eta_chi_q4_masses}
\end{figure*}

\begin{figure*}[htbp]
	\centering
		\includegraphics[width=0.3\textwidth]{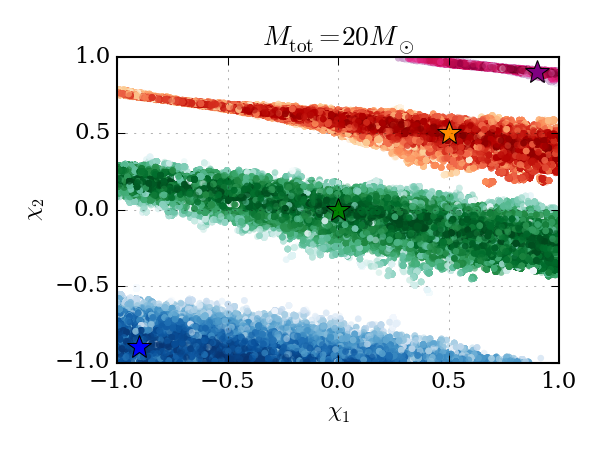}
		\includegraphics[width=0.3\textwidth]{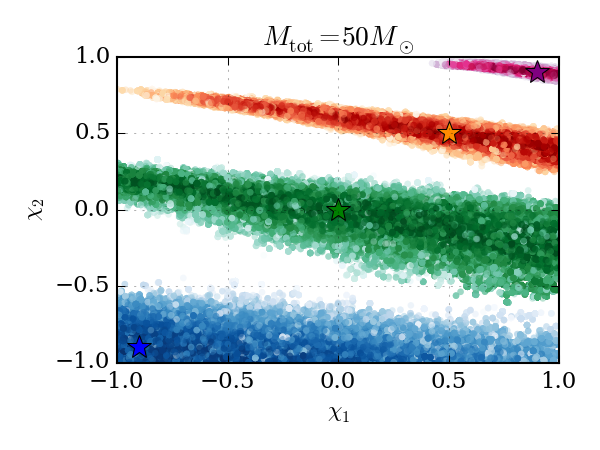}
		\includegraphics[width=0.3\textwidth]{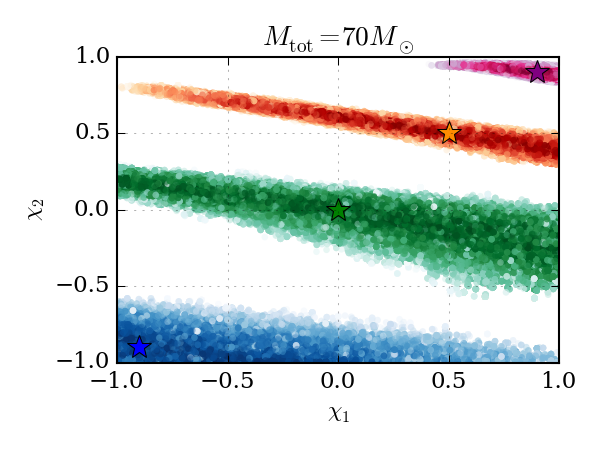}
		\includegraphics[width=0.3\textwidth]{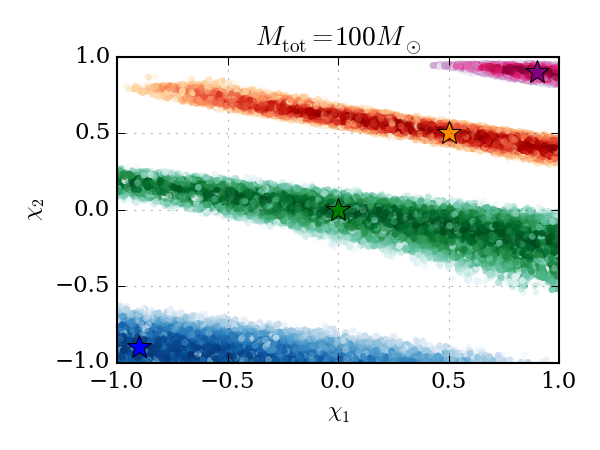}
		\includegraphics[width=0.3\textwidth]{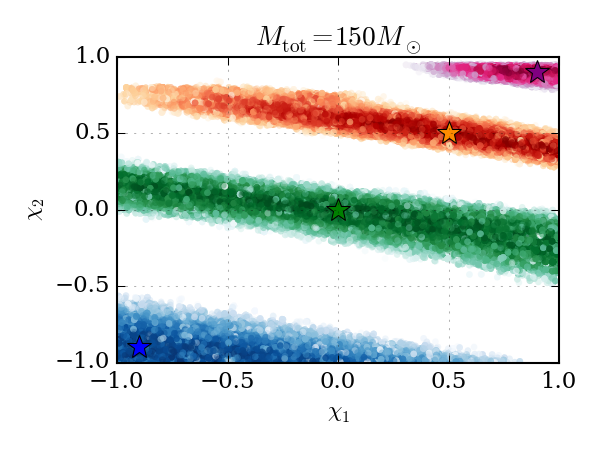}
		\includegraphics[width=0.3\textwidth]{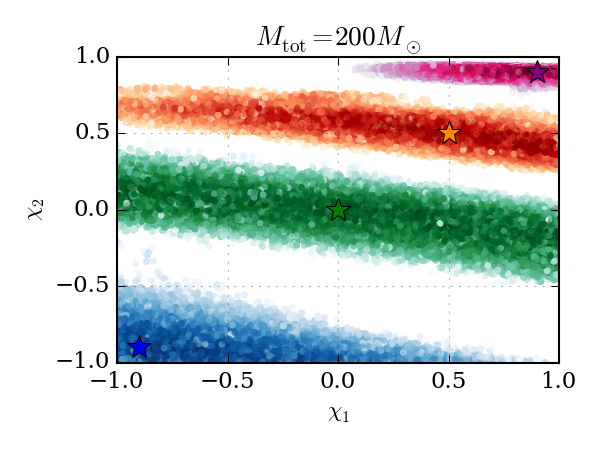}
	\caption{$\chi_1$-$\chi_2$ posteriors for mass-ratio $q=4$ and a range of total masses.
	}
	\label{fig:chi1_chi2_q4_masses}
\end{figure*}

\begin{figure*}[htbp]
	\centering
		\includegraphics[width=0.45\textwidth]{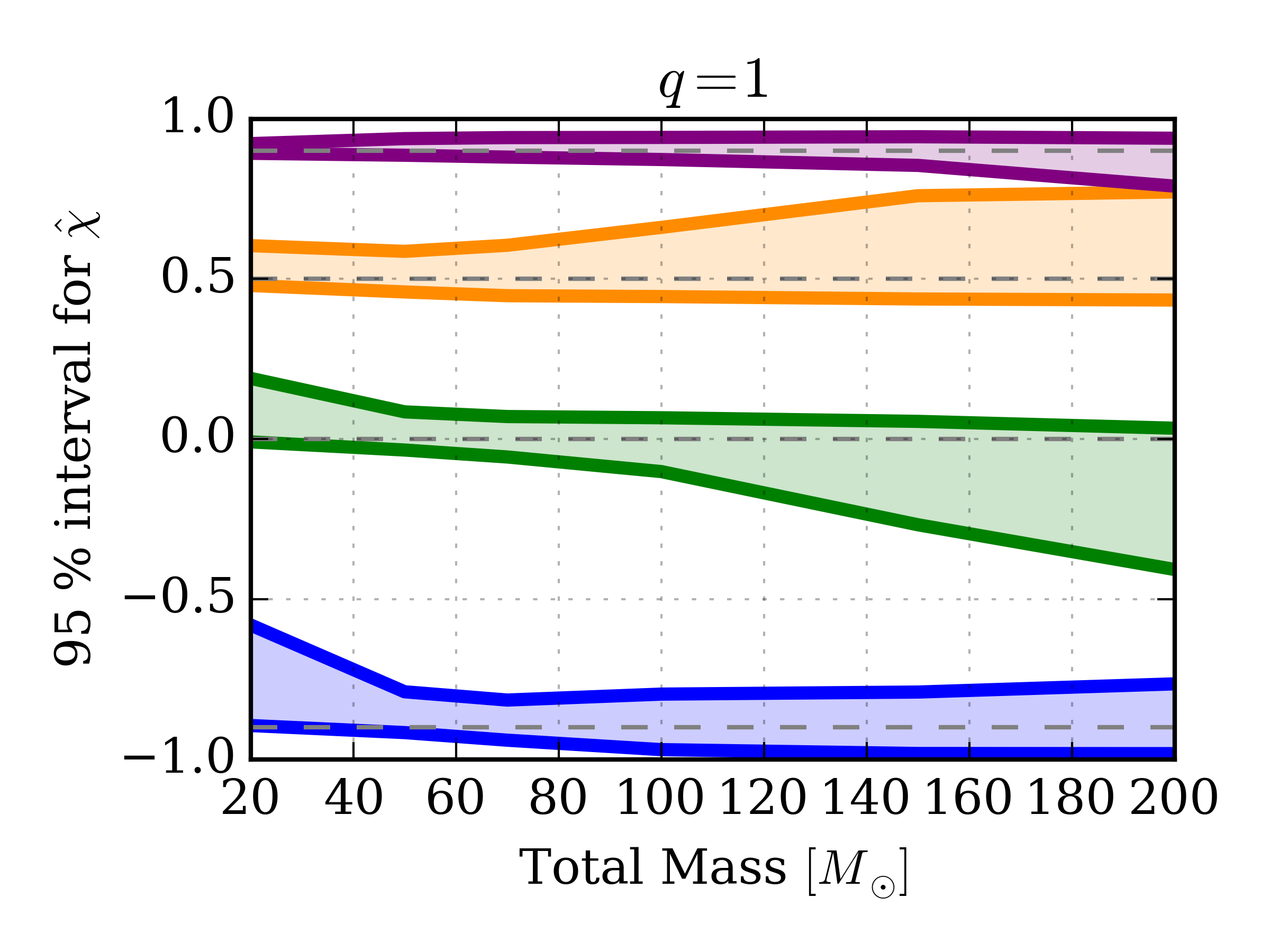}
		\includegraphics[width=0.45\textwidth]{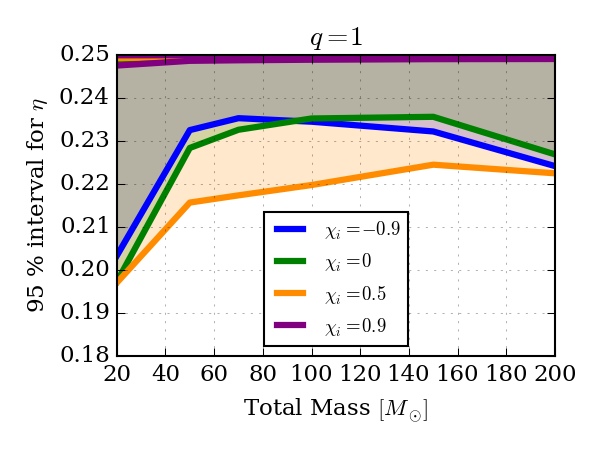}
		\includegraphics[width=0.45\textwidth]{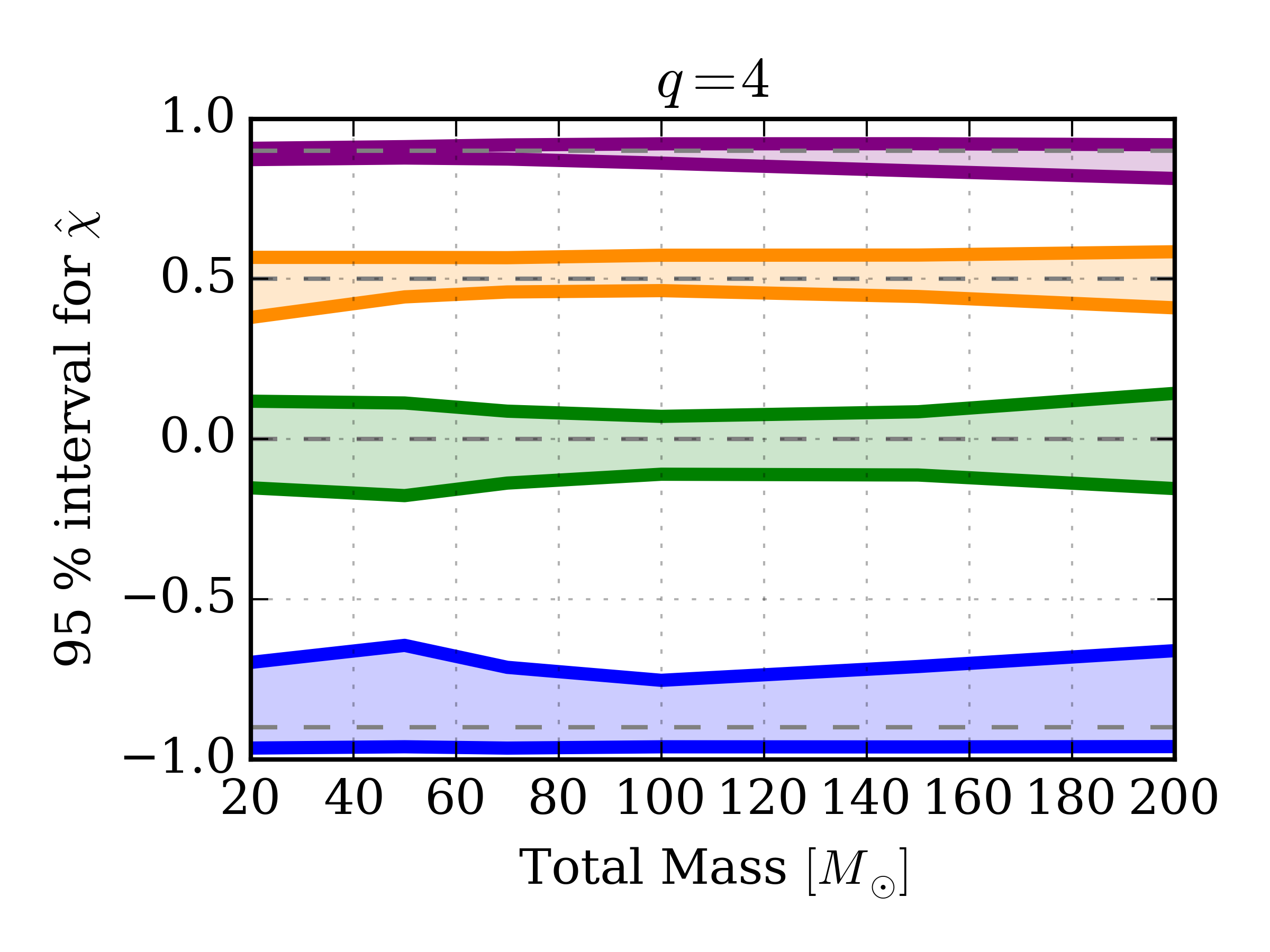}
		\includegraphics[width=0.45\textwidth]{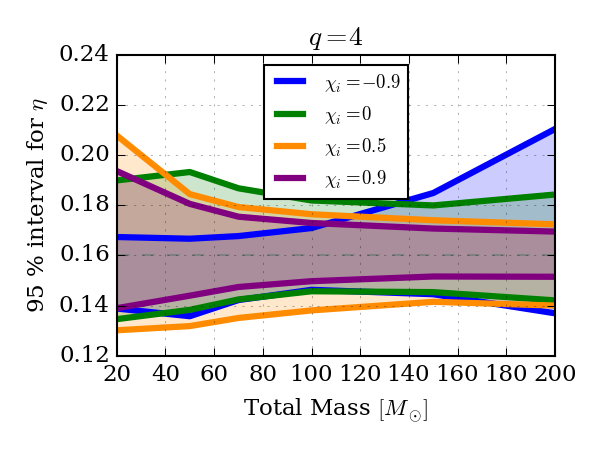}
		\includegraphics[width=0.45\textwidth]{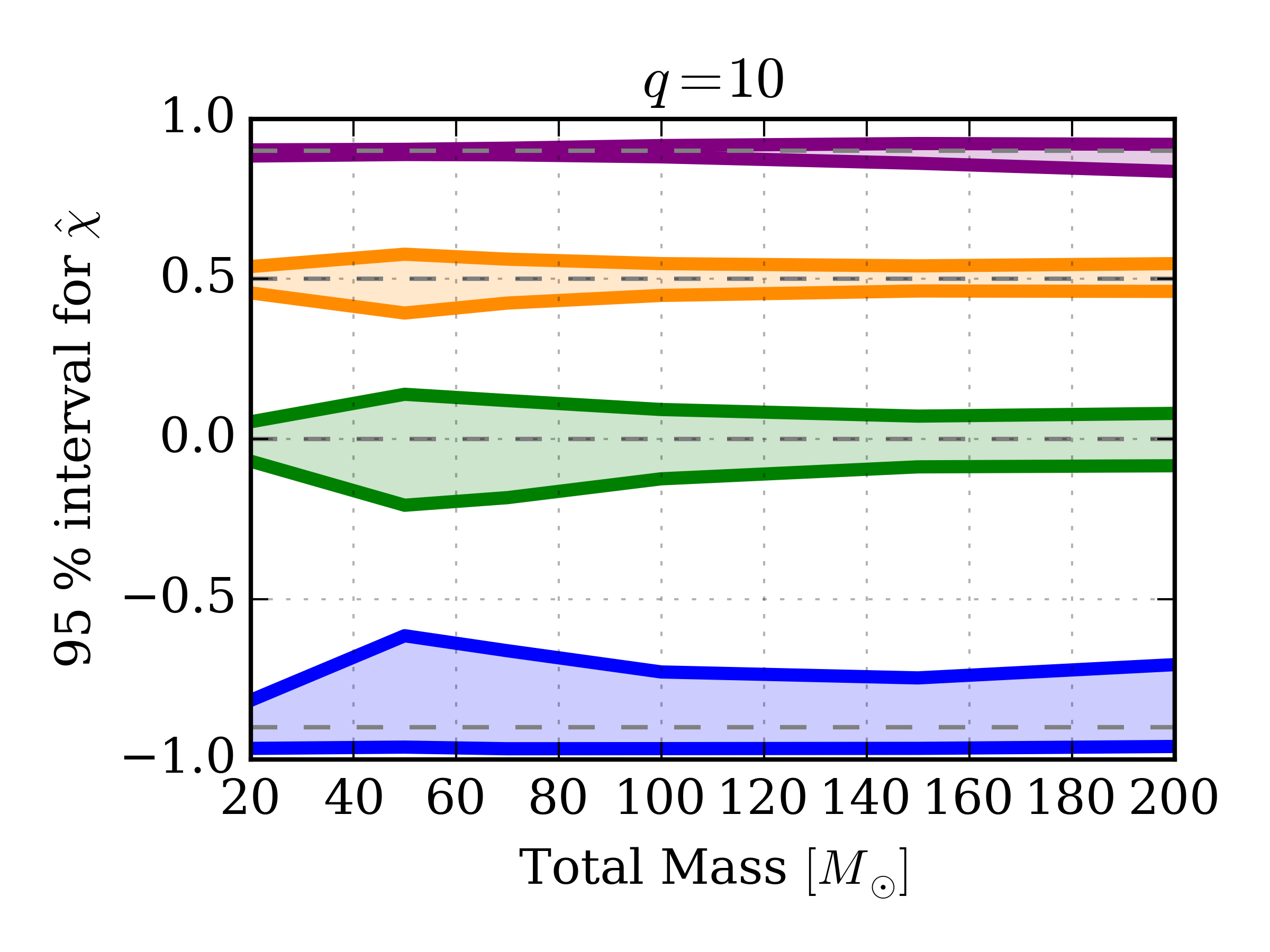}
		\includegraphics[width=0.45\textwidth]{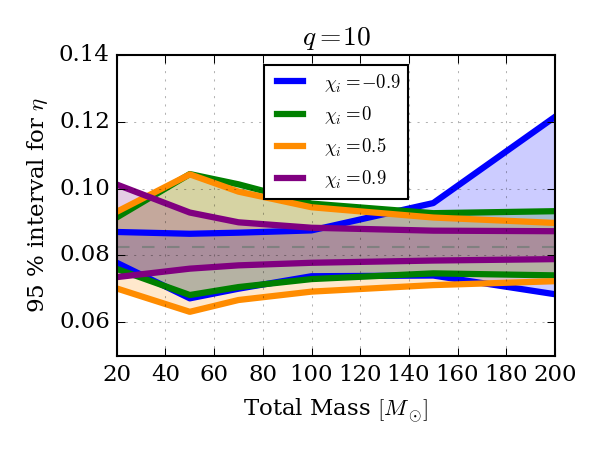}
	\caption{$95 \%$ intervals for the rescaled reduced spin $\hat\chi$ (left) and the symmetric mass-ratio $\eta$ (right) as a function of total mass for a series of configurations at mass-ratios $q=1,4,10$ top to bottom.
	}
	\label{fig:chi_eta_95pc_CI_q4}
\end{figure*}

One of the motivations for this study was to explore to what extent the approximate reduced-spin degeneracy is
affected when the \GW detectors are sensitive to comparable power from both the
inspiral and the merger-ringdown. 
This relative distribution of power is a function of the total mass of the system. In this section we investigate how
 the picture given in Sec.~\ref{sub:equal_spin} depends on the total mass, and comment on the measurement 
 accuracy of the mass-ratio, reduced spin and component spins. The accuracy of recovery for chirp mass and total 
 mass was studied in Refs.~\cite{Veitch:2015ela, Graff:2015bba, Haster:2015cnn}.

Fig.~\ref{fig:chi1_chi2_q4_masses} shows how the spin posteriors change with total mass. It is apparent that the posteriors do not significantly qualitatively change.
Recent studies have pointed out that SEOBNRv2 is not expected to be accurate for high aligned spin systems at unequal mass~\cite{Khan:2015jqa,Prayush-Chu-SEOBNR-Phenom-comparison}. Therefore our results for $\chi_i = 0.9$ might change once an improved model is available.

In Fig.~\ref{fig:eta_chi_q4_masses} we show how the mass-ratio--reduced spin posteriors change with total mass. 
The correlation is very strong at low mass, with the posterior samples approximately following a straight line. 
As the mass increases the regions become more fuzzy and thicker, indicating a weaker correlation. This implies that 
measurement accuracy for $\chi$ decreases slightly if the posterior is almost parallel with the $\eta$ axis, i.e., for high
 aligned spins. If the spin is smaller, or the spins anti-aligned, the slope is large enough that the increased width of the 
 region does not greatly affect the $\chi$ measurement. 
 
 To condense the information further the left panel of Fig~\ref{fig:chi_eta_95pc_CI_q4} shows the $95 \%$ credible 
 interval for $\chi$ for mass-ratios $q=1, 4, 10$. There is some broadening of the intervals for equal-mass systems at 
 high total mass, but overall the reduced spin is pretty consistently well measured for a wide range of the parameter 
 space. It is exceptionally well measured for high aligned spins.

The measurement accuracy of the symmetric mass-ratio $\eta$ depends both on the total mass and on the spin. 
For high aligned spins the accuracy actually improves as the total mass is increased. For high anti-aligned spins the 
accuracy first improves with the total mass, and is best around $M_\text{tot} \sim 70$ -- $100 M\odot$. It then worsens
 considerably if the mass is further increased. This becomes very obvious in the right panel of Fig~\ref{fig:chi_eta_95pc_CI_q4} for unequal mass systems. The behavior is reversed at equal mass.


\subsection{Dependence on SNR} 
\label{sub:dependence_on_snr}

\begin{figure*}[htbp]
	\centering
		\includegraphics[width=0.45\textwidth]{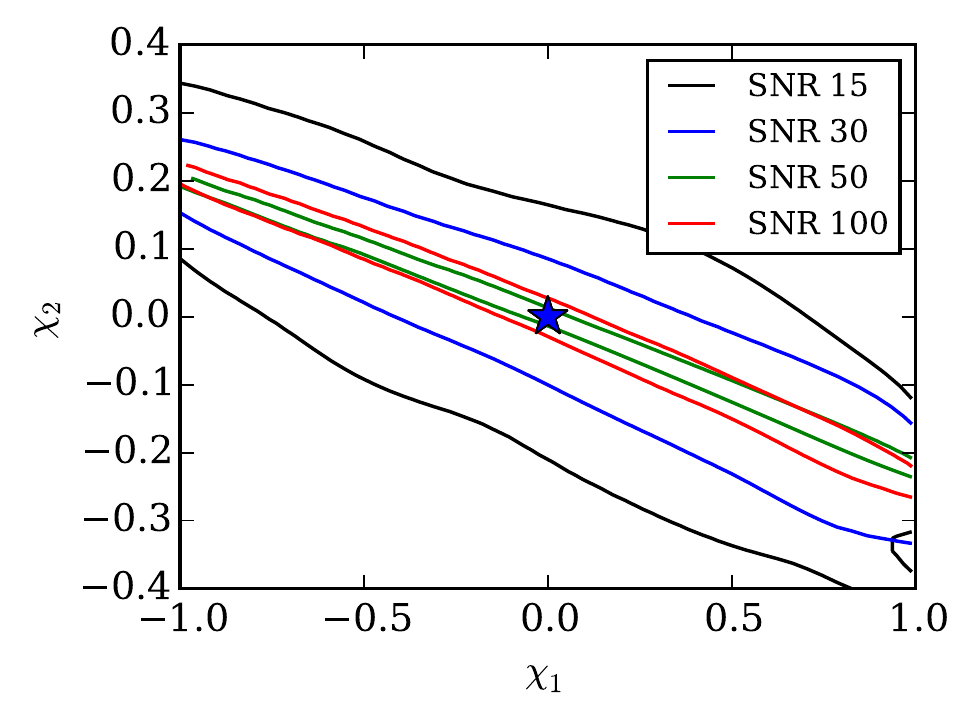}
		\includegraphics[width=0.45\textwidth]{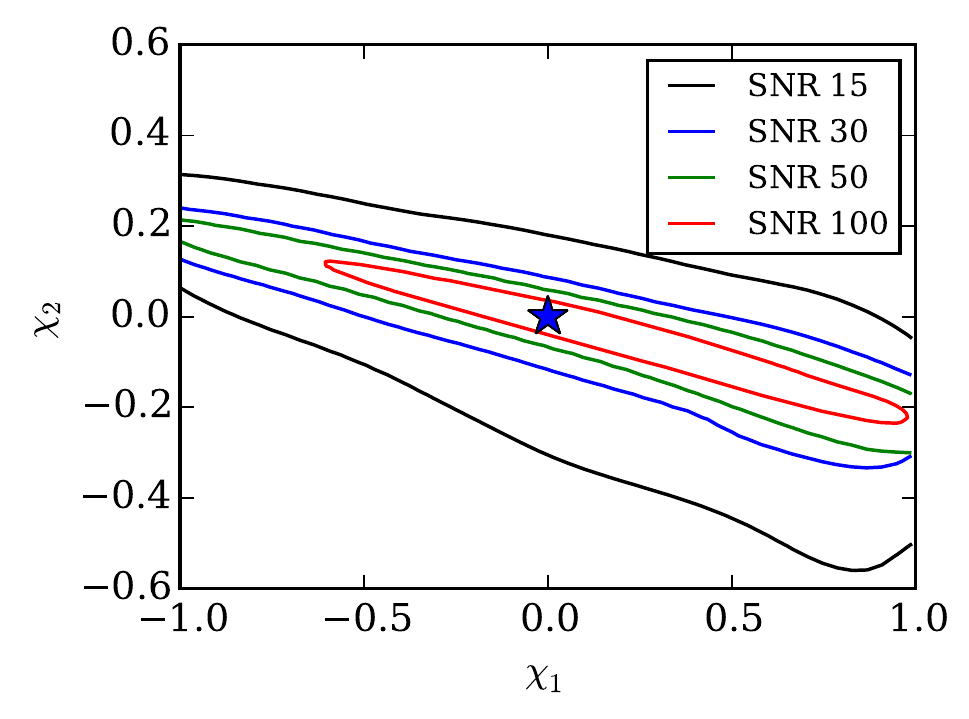}
	\caption{$95\%$ joint credible regions of the component spins for a
$q=4$ non-spinning system at total mass of $12 M_\odot$ (left) and $100 M_\odot$
(right) as a function of \SNR.
	}
	\label{fig:chi1_chi2_95pc_CRs_SNR_q4a0a0}
\end{figure*}

\begin{figure*}[htbp]
	\centering
		\includegraphics[width=0.3\textwidth]{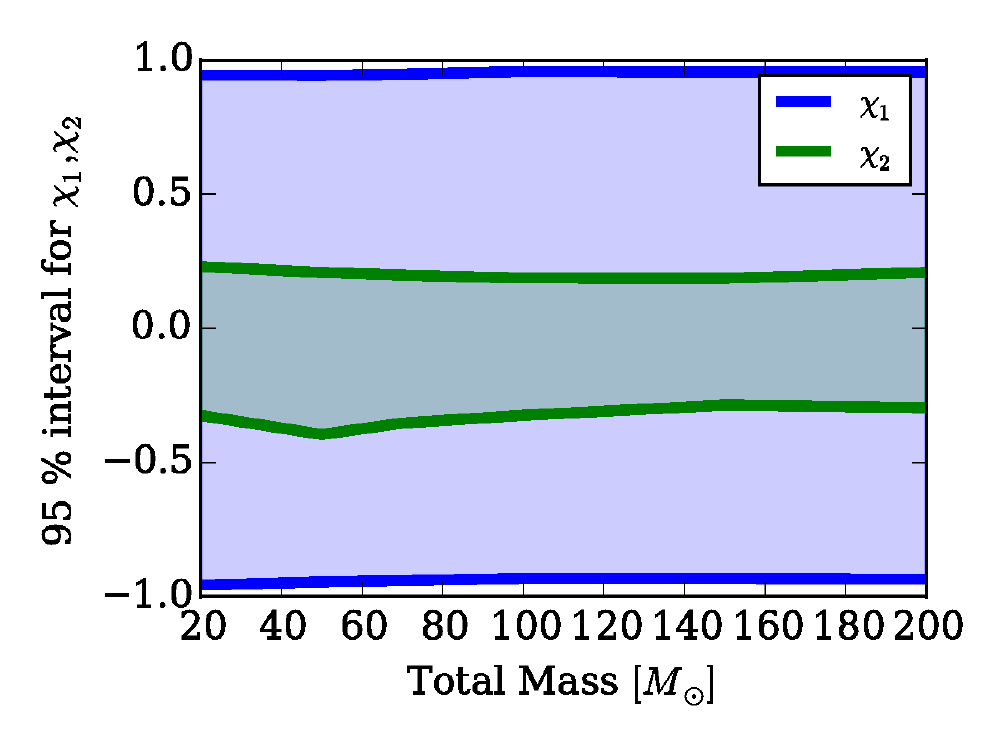}
    \includegraphics[width=0.3\textwidth]{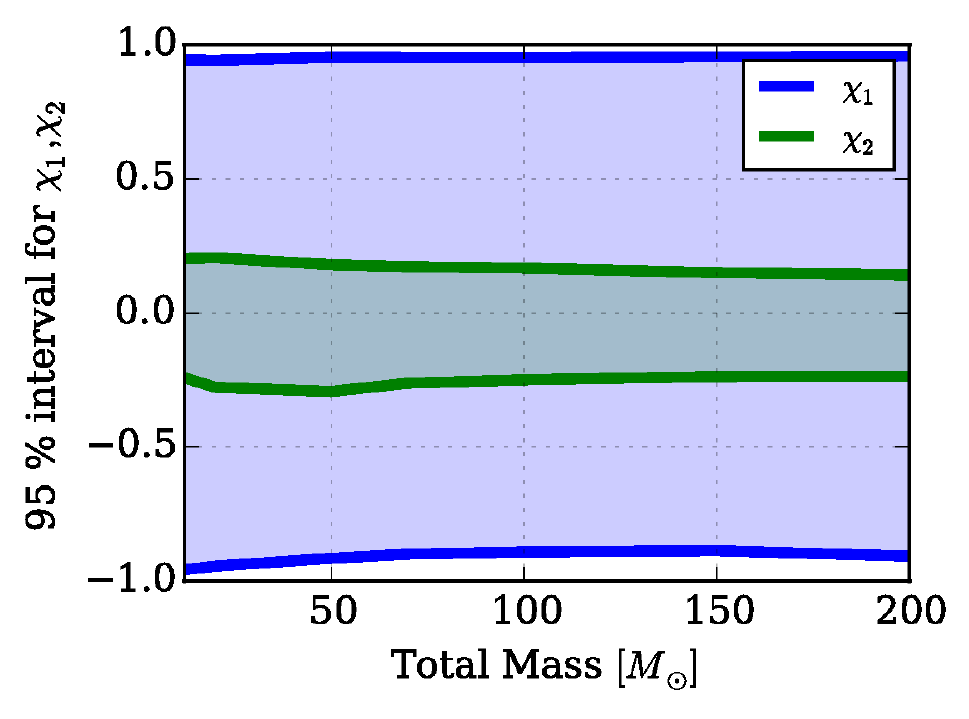}
		\includegraphics[width=0.3\textwidth]{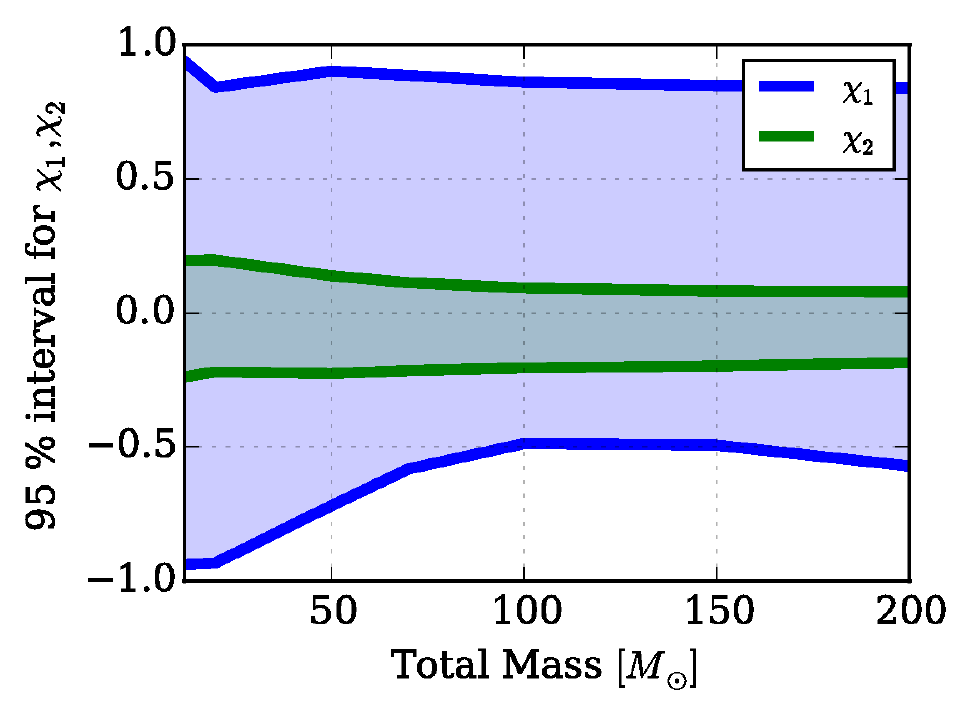}
	\caption{$95\%$ credible intervals for the component spins for
non-spinnning binaries at mass-ratio $q=4$ as a function of total mass. The
panels show the spin on the small black hole (blue) and on the large black hole
(green) for \SNR 30 (left), \SNR 60 (middle), and \SNR 100 (right).
	}
	\label{fig:chi_bands_q4_SNRs}
\end{figure*}

\begin{figure}[htbp]
  \centering
    \includegraphics[width=.45\textwidth]{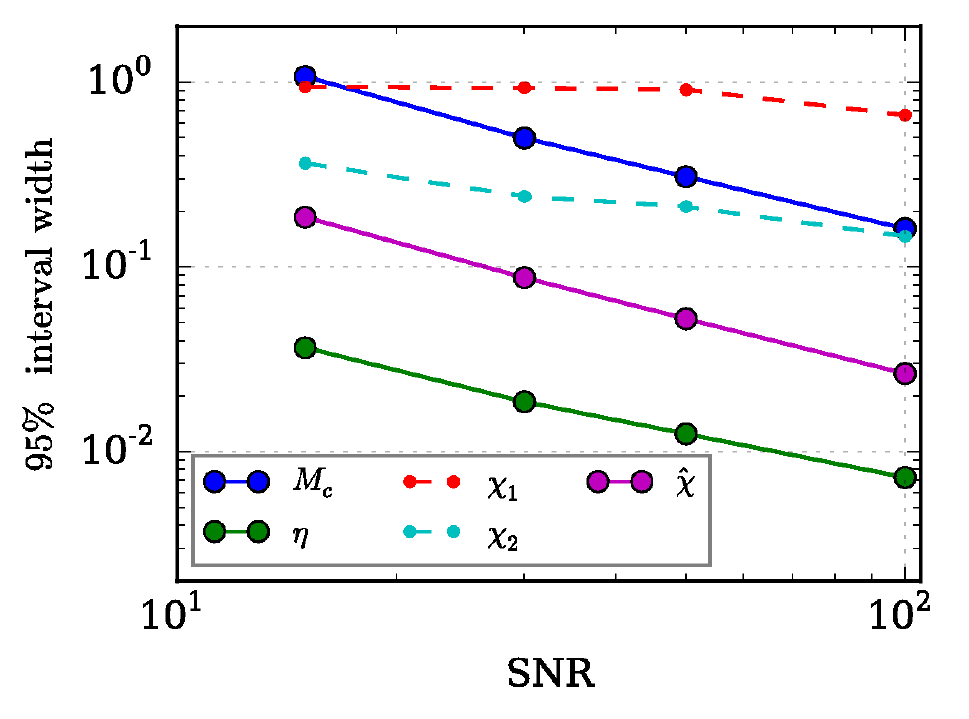}
  \caption{The width of $95 \%$ credible intervals for chirp mass $M_c$,
symmetric mass-ratio $\eta$, rescaled reduced spin $\hat\chi$, and the aligned
spin components $\chi_1, \chi_2$ against \SNR. As in
Fig.~\ref{fig:chi1_chi2_95pc_CRs_SNR_q4a0a0} results are shown for a
non-spinning system at mass-ratio $q=4$ at a total mass of $100 M_\odot$.}
  \label{fig:95pc_CI_SNR_q4a0a0_Mtot100}
\end{figure}

So far we have discussed results at a fixed \SNR of 30. If the \SNR is not too
low we can expect that the parameter measurement uncertainties scale inversely
with the \SNR~\cite{Vallisneri:2007ev}.
The mass-ratio---spin degeneracy is not absolute, and at sufficient \SNR we
expect to be able to measure the spin of
the smaller black hole. How high must the \SNR be? We consider a non-spinning
system to avoid the $\chi_2$ extreme-Kerr
 boundary. We choose a mass-ratio $q=4$. At \SNR 30 we cannot constrain the spin
on the small black hole for this system, 
 irrespective of the total mass.

We know from \PN results that \acp{SNR} of hundreds to thousands are
required to constrain the spin on the small black hole
\cite{Cutler:1994ys,Poisson:1995ef,Nielsen:2012sb,Ohme:2013nsa}. 
By including merger-ringdown we expect that we can constrain this spin at a
lower \SNR if the total mass is chosen so 
that inspiral and merger-ringdown both contribute significantly.
Fig~\ref{fig:chi1_chi2_95pc_CRs_SNR_q4a0a0} shows that this is indeed the case. At a total mass of $100 M_\odot$ 
the spin on the small black hole, $\chi_1$, can be somewhat constrained at \SNR
100.
Fig.~\ref{fig:chi_bands_q4_SNRs} illustrates how these results depend on the
total mass. At \SNR 30 (left) the spin on the small black hole, $\chi_1$, cannot
be constrained at all, at \SNR 60 (middle) the lower bound starts to move away
from $-1$, while at \SNR 100 (right) we see that the measurement accuracy for
$\chi_1$ improves noticeably with the total mass until about $100 M_\odot$ and
stays roughly the same until $200 M_\odot$.

Systematic errors in the ROM and the underlying EOB model become important at
high \acp{SNR}. In the region spanned by the posterior in
Fig~\ref{fig:chi1_chi2_95pc_CRs_SNR_q4a0a0} the ROM is only 
indistinguishable~\cite{Lindblom:2008cm} from the original SEOBNRv2 waveform for
\acp{SNR} of order $\sim$22 for the worst mismatch against SEOBNRv2. With
respect to the median ROM error this improves to an \SNR of $\sim$55. Parameter
estimation at 
\SNR 100 would require much more accurate models. However, even SEOBNRv2 is no
longer indistinguishable 
against its NR calibration waveforms at such a high \SNR. 
The indistinguishability criterion is sufficient, but not necessary and it has been shown that it is in practise far too conservative where parameter estimation is concerned~\cite{Littenberg:2012uj}.
While these results should be taken with a grain of salt we want to make the point that if the ROM is a reasonably 
smooth model with respect to parameter variations, it can still be used even to
give an estimate of the qualitative behaviour of measurements at high \SNR. 
We can sanity check the results in two ways: 
At low mass the credible regions obtained from the ROM agree with regions from IMRPhenomD, which fully incorporates two-spin
effects during the early inspiral~\cite{Khan:2015jqa}, and the credible regions scale as expected with 
\SNR (see Fig.~\ref{fig:95pc_CI_SNR_q4a0a0_Mtot100} and Fig 7
of~\cite{Veitch:2015ela}).

These results suggest that the combined information from the inspiral and the 
merger-ringdown \emph{does} indeed improve the measurement of the small black
hole's spin, but this effect only 
becomes significant at \acp{SNR} of 100 or beyond. For such signals, a two-spin
model can begin to place constraints 
on the spin of the smaller black hole, beyond what would already be known from a
single-spin model and the consequences 
of the Kerr limit. For lower \acp{SNR}, however, we do not seem to recover
significant extra information from the two-spin 
model. Note also that the improvement in the spin measurement is not strong --- even for the strongest signals we
expect to be able to detect with \aLIGO at design sensitivity, we do not expect
to be able to constrain the small black hole's
spin to anything other than a statement that it is ``small'' or ``large''.


\section{Discussion} 
\label{sec:discussion}

We have examined our ability to measure individual black-hole spins in \aLIGO
and \AdV observations of aligned-spin
black-hole-binary mergers. Our results are based on the SEOBNRv2 model, which aims to capture the effect of both 
black-hole spins on the \GW signal. This model includes only the dominant
$(\ell=2,|m|=2)$ harmonics, and \emph{does
not} model precession effects. We will make some comments below on how we expect these approximations to affect
our overall conclusions. 

For low-mass binaries (less than $\sim$10$\,M_\odot$), we expect a degeneracy
between the binary's mass ratio and a 
combination of the spins, and another degeneracy between the two spins
themselves, to make it difficult to measure the
individual spins (in particular the spin of the small black hole) below
\acp{SNR} of
$\mathcal O(1000)$. For higher masses, where the merger and ringdown 
contribute increasingly to the overall \SNR, an alternative degeneracy (between
systems with the same final mass
and spin) dominates, and it is conceivable that observations that include both
the inspiral and merger-ringdown will allow us to 
break these degeneracies and constrain measurements of both spins. That is what we have investigated here. 

For the configurations that we have studied, this does not appear to happen
until we reach \acp{SNR} of $\sim$100. Below
that the spins are constrained only by our knowledge of the extreme Kerr limit. This is illustrated in Fig.~\ref{fig:eta_chi_chi1_chi2_posteriors_Mtot50Msun}. For some
configurations our ability to measure both spins is indeed optimal at masses where both the inspiral and merger-ringdown
contribute comparably to the total \SNR (see Fig.~\ref{fig:chi_bands_q4_SNRs}),
at masses of $\sim$100$\,M_\odot$, 
but this effect will only aid
measurements of signals with very high \SNR (see
Fig.~\ref{fig:chi1_chi2_95pc_CRs_SNR_q4a0a0}). We remind the reader that, 
assuming a uniform volume distribution of 
sources in the universe, and a threshold detection \SNR of $\sim$10, \acp{SNR}
above 100 will occur in roughy one in a thousand
observations. The most optimistic current rate estimates~\cite{Abadie:2010cf,Dominik:2014yma} suggest that there might be several 
such observations
per year when \aLIGO and \AdV reach their design sensitivity, but in these cases
we would only be able to constrain the
smallest black hole (or, in the case of equal-mass systems, both black holes) to
possessing either ``high'' or ``low'' spin.

Our study is not exhaustive, but we have considered configurations over a wide sampling of the parameter space (up to 
mass ratios of 1:10). The results could change depending on the overall systematic accuracy of the SEOBNRv2 
model. The model was found to be less accurate for mixed aligned / anti-aligned configurations~\cite{Prayush-Chu-SEOBNR-Phenom-comparison} up to mass-ratio 1:3. Beyond these mass-ratios the model has not been systematically checked against unequal-spin waveforms.
There are also known inaccuracies for high-aligned-spin systems near merger~\cite{Khan:2015jqa,Prayush-Chu-SEOBNR-Phenom-comparison}.
However, over much of the parameter space that we have considered, we expect the model to be robust, and we do not
expect the qualitative picture to change.

\begin{figure*}[h]
  \centering
\includegraphics[width=.45\textwidth]{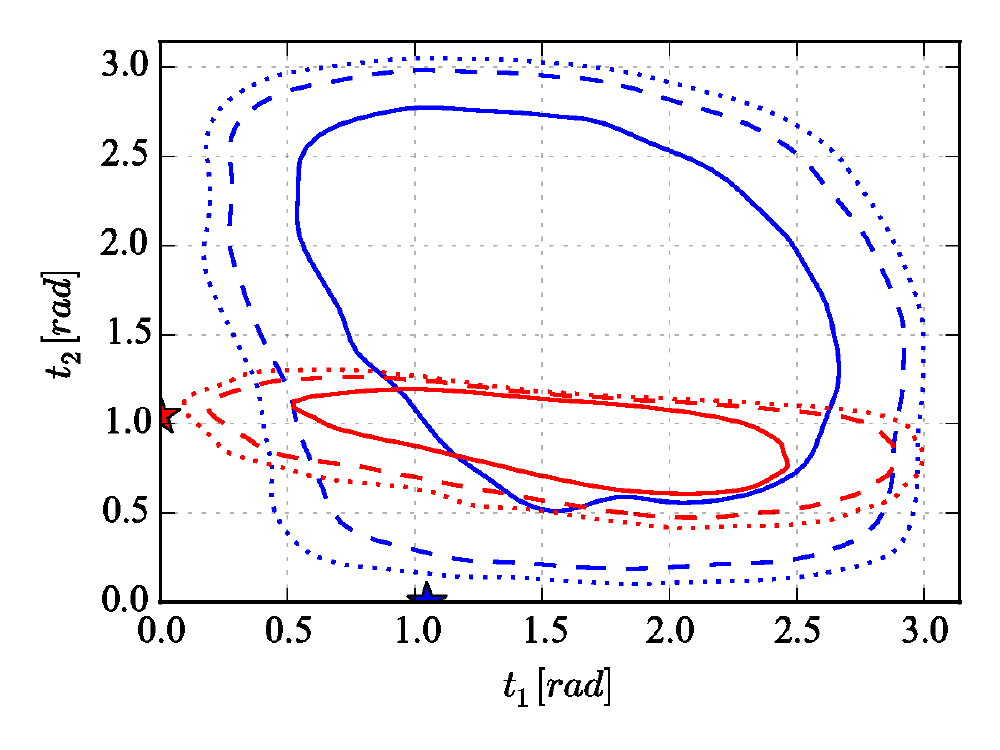}
    \includegraphics[width=.45\textwidth]{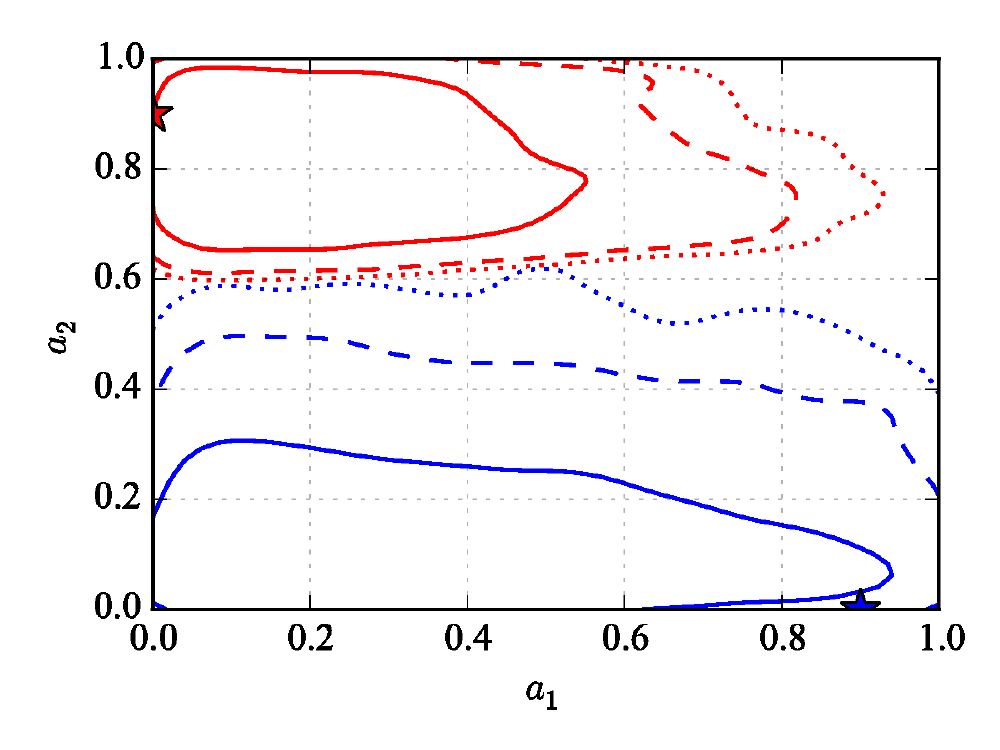}

  \caption{$68, 90, 95 \%$ credible regions for the spin tilt angles (left) and
spin magnitudes (right) for two configurations at $q=3$ and $M_\mathrm{tot} = 12
M_\odot$. We show CRs for precessing spin on the small black hole (blue) or the
large black hole (red). The true values of the spin magnitudes $0.9$ and tilt
angles $\pi/3$ are indicated by stars.}
  \label{fig:STT4_precessing_Q3_CR}
\end{figure*}

\begin{figure*}[h]
  \centering
    \includegraphics[width=.45\textwidth]{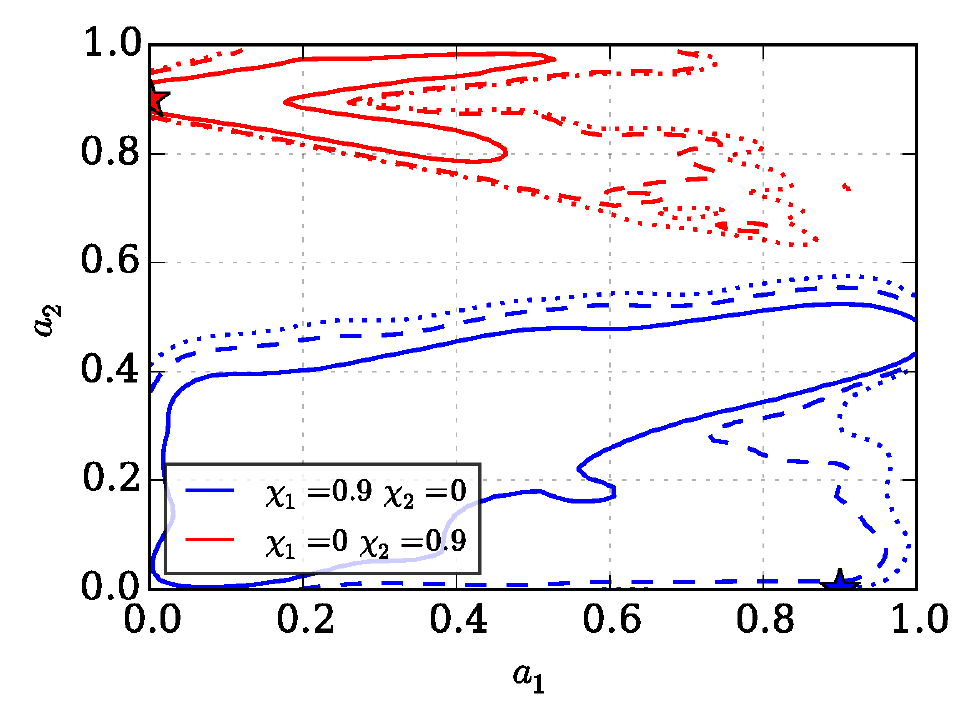}
    \includegraphics[width=.45\textwidth]{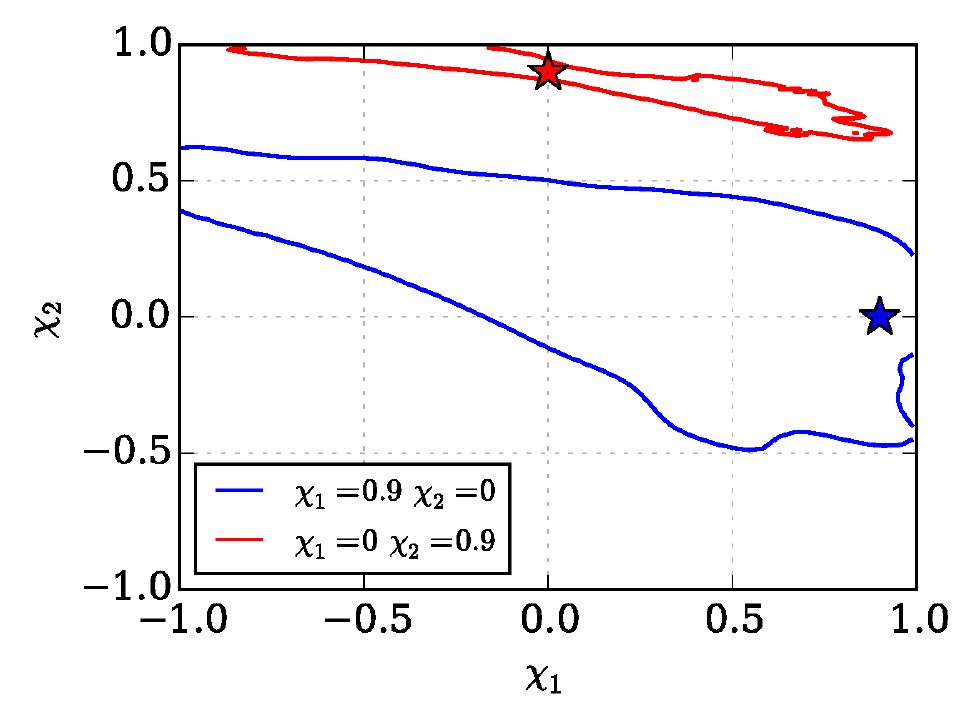}
  \caption{$68, 90, 95 \%$ credible regions for aligned-spin SEOBNRv2-ROM
configurations with $q=3$ and spin $0.9$ on the small black hole (blue) or the
large black hole (red). The left panel shows the spin magnitudes. The peculiar
shape of the regions can be explained by the elongated regions in the component
spins (right panel).}
  \label{fig:aligned-q3-comparison-cases-to-STT4_precessing_Q3_CR}
\end{figure*}

The most important effect that we have not included in this study is precession. In configurations where the binary's
total angular momentum is highly inclined with respect to the detector, strong
modulations of the \GW frequency and amplitude may
be detectable. These add more structure to the waveforms. Some studies have
shown how the inclusion of precession
can improve measurements of the mass ratio and the spin of the larger black 
hole~\cite{Chatziioannou:2014coa,O'Shaughnessy:2014dka,Littenberg:2015tpa,Farr:2015lna}, although there are still indications
that it is only the spin of the larger black hole that can be measured with any reasonable accuracy~\cite{Vitale:2014mka,Schmidt:2014iyl}.
To date no studies have explicitly considered our ability to measure individual spins, or have considered the effects of merger and
ringdown. For such systems, we note that we
will observe only a small number of precession cycles, so the effect of precession on our measurements may be small.  
It should also be borne in mind that, if we measure the binary's parameters with a model with a larger number of physical 
parameters, we may well \emph{increase} the size of the credible regions for each individual parameter; in this sense, the inclusion
of precession may in some cases \emph{worsen} our spin measurements. 

As an illustration of the effects of precession on individual spin measurements, we consider an inspiral-only example, for which 
\PN two-spin precessing models are available.
We study two configurations with mass-ratio 1:3, total mass of $12 M_\odot$, dimensionless 
spin magnitude $0.9$ on either the small or the large black hole, with the other
black hole non-spinning.
The spin tilt angle is $t_i = \arccos \left(\hat L \cdot \hat S_i \right) =
\pi/3$, where $i$ refers
the spinning black hole.
The binaries are viewed under an inclination of $3\pi/4$, at which we expect precession effects to be strong. 
We use SpinTaylorT4~\cite{LAL-web} without amplitude corrections both for the signal and for recovery with a lower frequency cutoff of 
$f_\text{low} = 40 \text{Hz}$ at \SNR $30$ and a sampling rate of $4096
\text{Hz}$. 
The simulations were performed with the nested sampling code from \texttt{lalinference}~\cite{Veitch:2014wba} for a three-detector 
\aLIGO-\AdV setup.

In Fig.~\ref{fig:STT4_precessing_Q3_CR} we compare the recovery of the spin magnitude and the tilt angles for the two configurations. 
If the spin is on the large black hole its tilt angle can be recovered fairly
accurately, while the tilt angle cannot be constrained if the spin is on 
the small black hole. The tilt angle of the non-spinning companion is not
well-defined for the signal and cannot be recovered. 
The spin magnitude can be recovered with some accuracy only if the larger black
hole is spinning. Similar behaviour has been seen for other configurations
in~\cite{Vitale:2014mka,SVcomment}.

Aligned spin comparison cases using SEOBNRv2\_ROM are shown in 
Fig.~\ref{fig:aligned-q3-comparison-cases-to-STT4_precessing_Q3_CR}. These configurations use the same setup as the 
precessing simulations, in particular also $f_\text{low} = 40 \text{Hz}$, except
that the black holes carry aligned spin of $+0.9$ on 
the small or large black hole. The regions are elongated along the constant
effective spin direction and by taking the modulus of the 
component spins the regions become V-shaped. We would expect to see a similar effect if we used a precessing-binary model
in the parameter estimation, if we made the additional requirement that the tilt angles had to be close to either 0 or $\pi$. 
Apart from this notch the regions cover comparable area in the spin magnitudes 
as shown in Fig.~\ref{fig:STT4_precessing_Q3_CR} (right). The measurement of the individual spins has not improved at all
in the precessing case. 

Other effects we do not consider in this study are spherical harmonic
modes beyond the dominant $(\ell=2, \vert m \vert = 2)$ contributions. Higher
harmonics become more important as the mass ratio is 
increased~\cite{Littenberg:2012uj,Graff:2015bba,Varma:2014jxa,
Wade:2013hoa,Capano:2013raa},  and also add more structure to 
the waveforms. We would expect the inclusion of higher harmonics to improve the measurement of individual spins, and we once
again have the possibility that there are ideal configurations (and binary orientations) that make it possible to accurately measure
both spins. However, since in the present study we find that individual spins are very poorly constrained, and in most configurations 
the higher harmonics contribute less than a few percent of the total signal power, it seems unlikely that the situation will change dramatically. 
To address this question conclusively, of course, would require a two-spin higher-mode IMR model.

It is clear, then, that we require IMR models that include higher harmonics and precession in order to fully understand our
potential to measure individual spins with Advanced-detector observations. Even for models that include only the 
$(\ell=2, |m|=2)$ modes, we require far greater accuracy than currently available in order to fully quantify the accuracy of 
measurements from observations with \acp{SNR} greater than $\sim$30. These
qualifications aside, we expect that in general
individual spins will be measurable only in cases where the spins are both near-extremal, and (at least for aligned-spin
binaries), both aligned or both anti-aligned with the orbital angular momentum. 


\section{Acknowledgements} 

We thank John Veitch, Alex Nielsen, Vivien Raymond, Harald Pfeiffer, Prayush Kumar, Alessandra Buonanno, Guillaume Faye, Salvatore Vitale, Eric Poisson and Aaron Zimmerman for useful discussions and comments.
We thank Edward Fauchon-Jones for contributions at the early stages of the project.
MH was supported Science and Technology Facilities Council grant ST/H008438/1 and MP, MH and FO by ST/I001085/1.
\texttt{lalinference} simulations were carried out at Advanced Research Computing (ARCCA) at Cardiff.


\bibliography{spin}

\end{document}